\documentclass[twocolumn]{aastex62}
\pdfoutput=1 
\usepackage{amsmath,amstext}
\usepackage{apjfonts} 

\usepackage[hang,flushmargin]{footmisc} 
\usepackage{lipsum}
\usepackage{mwe}
\usepackage{rotating}
\usepackage{soul}

\usepackage[nameinlink,capitalize]{cleveref}
\crefname{section}{section}{section}
\Crefname{Section}{Section}{Section}

\usepackage{url}

\usepackage{graphbox,graphicx}
\usepackage[figure,figure*]{hypcap}
\usepackage[caption=false]{subfig}
\newcommand{\imr}{\texttt{IMRPhenomPv2}\xspace}
\newcommand{\msun}{M$_\odot$\xspace}

\newcommand{\cg}[1]{\textcolor{black}{#1}}

\graphicspath{{./figs/}}

\hypersetup{
    allcolors = {blue},
}

\bibpunct{(}{)}{;}{a}{}{,}

\def\msun{{\rm\,M_\odot}}

\definecolor{orange}{rgb}{.8,0.4,0}

\defcitealias{LP98}{LP98}
\defcitealias{CowpBerger15}{CB15}

\shorttitle{Standard Siren Cosmology in the Era of LIGO A+, the Rubin Observatory, and Beyond}
\shortauthors{Chen et al.}

\begin{document}

\title{A Program for Multi-Messenger Standard Siren Cosmology in the Era of LIGO A+, Rubin Observatory, and Beyond}

\newcommand{\LIGOlabMIT}{\affiliation{LIGO Laboratory, Massachusetts Institute of Technology, 185 Albany St, Cambridge, MA 02139, USA}}
\newcommand{\MKI}{\affiliation{Department of Physics and Kavli Institute for Astrophysics and Space Research, Massachusetts Institute of Technology, 77 Massachusetts Ave, Cambridge, MA 02139, USA}}
\newcommand{\BHI}{\affiliation{Black Hole Initiative, Harvard University, 20 Garden St, Cambridge, MA 02138, USA}}
\newcommand{\CARNEGIE}{\affiliation{The Observatories of the Carnegie Institution for Science, 813 Santa Barbara St, Pasadena, CA 91101, USA}}
\newcommand{\COLUMBIA}{\affiliation{Department of Physics, Columbia University, New York, USA}}
\newcommand{\CFA}{\affiliation{Center for Astrophysics | Harvard \& Smithsonian, 60 Garden St, Cambridge, MA 02138, USA}}

\author{Hsin-Yu Chen}\BHI \MKI \LIGOlabMIT
\email{himjiu@mit.edu}

\author{Philip S. Cowperthwaite}\CARNEGIE

\author{Brian D. Metzger}\COLUMBIA

\author{Edo Berger}\CFA

\begin{abstract}
The most promising variation of the standard siren technique combines gravitational-wave (GW) data for binary neutron star (BNS) mergers with redshift measurements enabled by their electromagnetic (EM) counterparts, to constrain cosmological parameters such as $H_0$, $\Omega_m$, and $w_0$.  Here we evaluate the near- and long-term prospects of multi-messenger cosmology in the era of future GW observatories: Advanced LIGO Plus (A+, 2025), Voyager-like detectors  (2030s), and Cosmic Explorer-like detectors (CE, 2035 and beyond). We show that the BNS horizon distance of $\approx 700$ Mpc for A+ is well-matched to the sensitivity of the Vera C.~Rubin Observatory (VRO) for kilonova detections. We find that one year of joint A+ and VRO observations will constrain the value of $H_0$ to percent-level precision, given a small investment of VRO time dedicated to target-of-opportunity GW follow-up. In the Voyager era, the BNS-kilonova observations begin to constrain $\Omega_m$ with an investment of a few percent of VRO time. With the larger BNS horizon distance in the Cosmic Explorer era, on-axis short gamma-ray bursts (SGRBs) and their afterglows (though accompanying only some of the  GW-detected mergers) supplant kilonovae as the most promising counterparts for redshift identification. We show that five years of joint observations with Cosmic Explorer-like facilities and a next-generation gamma-ray satellite with localization capabilities similar to that presently possible with {\it Swift} could constrain both $\Omega_m$ and $w_0$ to $15-20\%$.  We therefore advocate for a robust target-of-opportunity (ToO) program with VRO, and a wide-field gamma-ray satellite with improved sensitivity in the 2030s, to enable standard siren cosmology with next-generation gravitational wave facilities.  
\end{abstract}

\keywords{binaries: close -- catalogs -- gravitational waves -- stars:
  neutron -- surveys}

\section{Introduction}
\label{sec:intro}

Although the field of cosmology has advanced dramatically over the past several decades, a number of puzzles and tensions have been identified that may challenge the standard $\Lambda$CDM paradigm.  Among these are differences in the Hubble constant $H_0$ as measured from the cosmic microwave background by Planck \citep{2018arXiv180706209P} and as determined from nearby Type Ia SNe \citep{2019ApJ...876...85R}, a tension which has now formally reached $\gtrsim 4\sigma$ (\citealt{Riess19}; however, see \citealt{Freedman+19}). Similarly, a $5.3\sigma$ tension with the Planck $H_0$ value was recently found from lensed quasars \citep{Wong+20}.  Although this difference could in principle be related to unmodeled systematics, it could also hint at new physics (e.g. \citealt{Knox&Millea19}). Likewise, dark energy remains almost as much of a mystery as when it was first discovered, and probing its nature remains a primary goal of ongoing and future surveys such as the Dark Energy Spectroscopy Instrument (DESI; \citealt{DESI+16}), \cg{the Nancy Grace Roman Space Telescope (Roman Space Telescope; \citealt{Dore+19})}, CMS-S4 \citep{Abazajian+19},  and the Vera C.~Rubin Observatory (VRO; \citealt{Ivezic+19}).

Even with the advent of new cosmology experiments over the next decades, there is clearly a need for new and independent methods to probe the cosmic expansion history.  One promising approach is the ``standard siren'' technique \citep{Schutz1986,HolzHughes05,Nissanke+11,2018Natur.562..545C,Feeney+19}, in which gravitational waves (GW) from a source of known strength are used to obtain an absolute distance scale, which is then combined with redshift information obtained from an associated electromagnetic (EM) counterpart\footnote{Although GW events alone (without EM counterparts), or in conjunction with galaxy survey statistics, can also provide a probe the cosmic expansion history (e.g., ``dark sirens''; \citealt{Chen&Holz16,2018Natur.562..545C,SoaresSantos+19,Fishbach+19}), such techniques are not the focus of this paper.}.  In practice, the most promising standard siren sources with ground-based GW detectors are the mergers of binary neutron stars (BNS), or of a neutron star with a stellar-mass black hole (NS-BH), as these sources generate luminous GWs and can give rise to detectable EM emission (e.g., \citealt{Metzger&Berger12,Berger14}).

Prospects for this technique are promising, and have been validated by GW170817, the first BNS merger detected by Advanced LIGO-Virgo \citep{LIGOGW170817}  with an associated EM counterpart \citep{LIGOGW170817grb,LIGO+17CAPSTONE}.  In particular, the discovery of optical kilonova emission about 11 hours after the merger (\cg{e.g., \citealt{2017Sci...358.1556C,GW170817DECam,2017Natur.551...67P}}) enabled an immediate identification of the host galaxy, NGC4993.  The redshift of the host combined with the GW luminosity distance of GW170817 led to a measurement of $H_0 = 70^{+12}_{-8}$ km s$^{-1}$ Mpc$^{-1}$ \citep{LIGOH0,Guidorzi+17}.  This value is consistent with both the Type Ia SN and Planck values of $H_0$, but is not sufficient to arbitrate between them.  Additional joint GW-EM events are needed to reduce the statistical uncertainty.   

By the mid 2020s, the current LIGO facilities are expected to surpass the nominal design of the second generation GW detector and achieve a better sensitivity known as ``A+'' \citep{LIGOLocalization}.  With these second generation GW detectors, percent level uncertainty in $H_0$ could be achieved with $\sim 50$ joint GW-EM detections \citep{2018Natur.562..545C}. In the 2030s different possible GW detector upgrades are under current consideration, including an optimal upgrade of the LIGO facilities, known as ``Voyager'', and new facilities such as the Einstein Telescope (ET) and Cosmic Explorer (CE; \citealt{LIGOG3Ref}).  By extending the range of BNS detections beyond the local Hubble flow, the standard siren technique becomes sensitive to additional cosmological parameters.  With ET, for example, it is estimated that a few percent uncertainty in $\Omega_m$ and $w_0$ can be achieved with $\sim 10^3$ joint GW-EM detections \citep{2010CQGra..27u5006S,2011PhRvD..83b3005Z,2017PhRvD..95d4024C}. 

Despite this great promise, the {\it practical} path to achieving a large number of joint GW-EM detections is unclear and has not been fleshed out in great detail.  Given the allowed volumetric BNS merger rate inferred by present LIGO/Virgo observations of $\approx 80-810$ Gpc$^{-3}$ yr$^{-1}$ \citep{Abbott:2020gyp}, events as close as GW170817 may occur as infrequently as once per decade.  Otherwise similar sources at greater distance will be dimmer electromagnetically and, given the much larger number of galaxies per error region, will require a different approach to discovering the EM counterpart than most follow-up efforts employed for GW170817.  Furthermore, for events with poor sky localization, the kilonova emission will be much more challenging to discover, as evidenced by follow-up observations in Observing Run 3 \cg{\citep{Gomez+19,Hosseinzadeh+19,Andreoni+20,Vierira+20,Ackley:2020qkz,Antier:2020nuy}}. This motivates the use of target-of-opportunity observations with VRO, whose unparalleled survey speed and sensitivity would make it the ideal tool for GW-EM astronomy in the 2020s and beyond (e.g., \citealt{Margutti+18,Cowperthwaite+19,Cowperthwaite+19b}).  

Even so, in the Voyager and CE/ET eras, we expect that the BNS merger sensitivity range will challenge even the VRO's capabilities for kilonova detections. Fortunately, a small fraction of BNS mergers (those viewed roughly on-axis) are expected to be accompanied by a more luminous EM counterpart, namely SGRBs and their associated afterglows, which are detectable to redshifts of $z\gtrsim 1$ \citep{Berger14}.  To date, the vast majority of SGRB afterglow, host galaxy, and redshift identifications were enabled by the the {\it Neil Gehrels Swift} satellite, which is capable of detecting the early X-ray afterglow with a localization of a few arcseconds.  As we demonstrate here, this approach may become the method of choice for joint GW-EM detections at $z\gtrsim 0.5$.     

In this paper, we explore the prospects for standard siren cosmology in the era of LIGO A+ and for subsequent generations of GW detectors.  Distinct from past work on this topic, we focus not only on the precision achievable given a number of joint detections, but also on what is practical to achieve with EM follow-up given planned or conceivable ground- and space-based telescope facilities. Indeed, as we show, given the substantial commitment of EM resources that a serious standard siren program would entail, future planning for such a program is warranted now.  

The paper is structured as follows.  In \S\ref{sec:facilities} we describe our assumptions about the capabilities of existing and future/proposed GW and EM facilities, and define several scenarios that combine these facilities to jointly observe BNS mergers and obtain their redshifts.  In \S\ref{sec:simulations} we describe Monte Carlo simulations of a large sample of BNS mergers to address the achievable precision of cosmological parameters for each scenario.  In \S\ref{sec:results} we describe our simulation results. Finally, in \S\ref{sec:program} we synthesize our findings and use them to make specific recommendations for a long-term program of multi-messenger standard siren cosmology.

\section{GW and EM Facilities}
\label{sec:facilities}

\begin{figure}
	\centering
	\includegraphics[width=1.0\columnwidth]{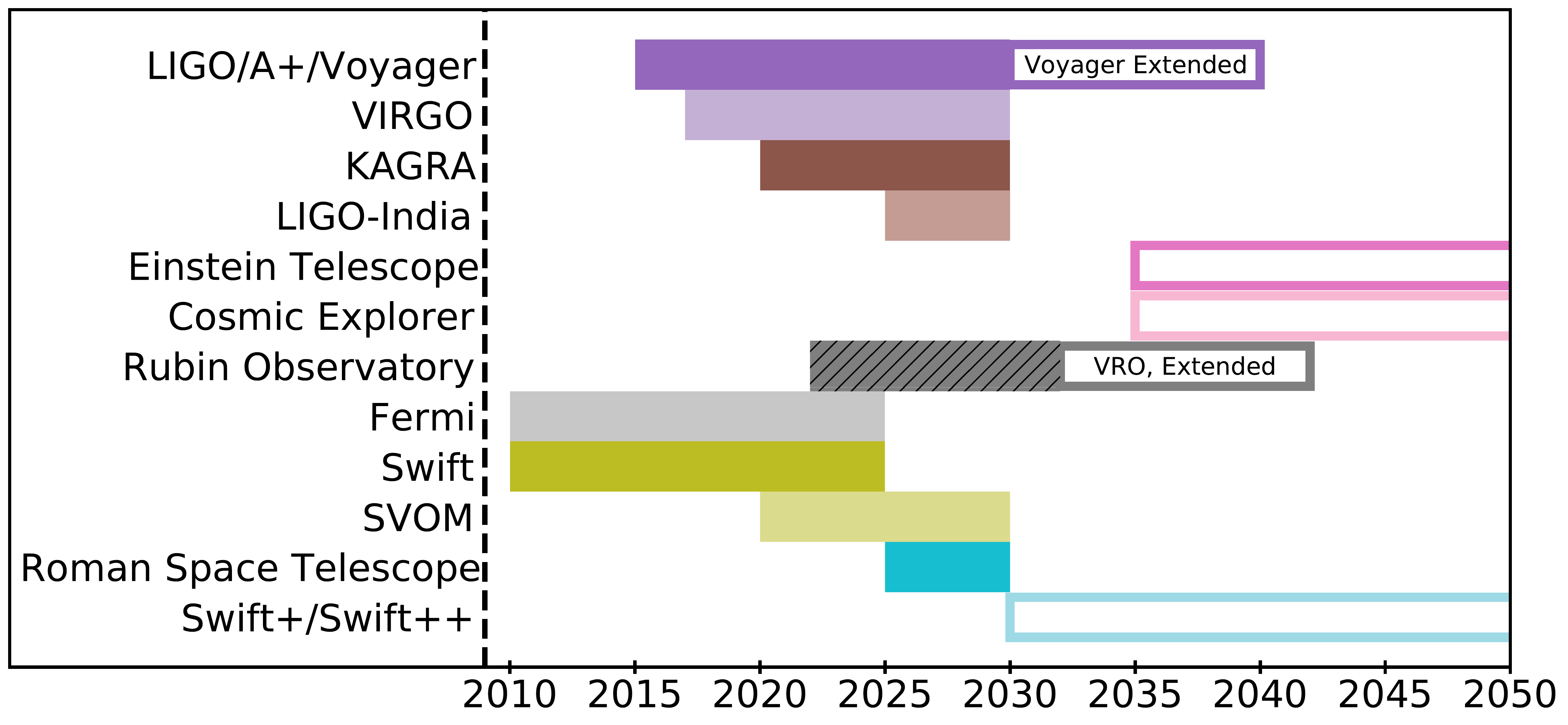}
	\caption{\label{fig:timeline}
Schematic timeline of existing (solid), funded (hatched), and proposed (open) GW and EM facilities over the next three decades.  {\it Swift+/Swift++} are hypothetical future gamma-ray satellites (see \S\ref{sec:facilities})}
\end{figure}

In Figure~\ref{fig:timeline} we show a rough schematic timeline of active, proposed, and envisioned GW and relevant EM facilities in the next three decades.  In this section we describe our assumptions about the reach of current or planned facilities, and define example programmatic choices regarding the GW events that could be followed up with a given EM technique.  As summarized in Table~\ref{tab:scenarios}, and discussed in detail below, we define several distinct ``scenarios'' that involve particular combinations of GW and EM facilities. In subsequent sections we assess how well each of these scenarios constrain cosmological parameters.  

We first consider the GW observatories.  We assume a network that includes LIGO-Hanford, LIGO-Livingston, and Virgo, upgraded from their capabilities at the end of the Advanced observing runs.  We use the projected A+, Voyager, and CE strain sensitivities from \url{https://dcc.ligo.org/LIGO-T1500293-v11/public}, and all three LIGO-Virgo detectors are assumed to operate with the same sensitivities \cg{(even in the CE era)}.  We assume 50\% duty cycle for simultaneous operation of all three detectors, motivated by the recent O1/O2 runs \citep{LIGOScientific:2018mvr}. The KAGRA \citep{KAGRA+19} and LIGO-India observatories are not directly included in our simulations~\footnote{Due to the duty cycle, detections made by four- or five detectors are less common.  Although we only include LIGO-Hanford, LIGO-Livingston, and Virgo in our simulations, we expect the results will be similar to the three-detector detections made in a five-detector network including KAGRA and LIGO-India.}.  Although the locations, numbers, and sensitivities of future detectors are still not known, our basic scenario provides a reasonable approximation of the expected landscape. 

On the EM side, we consider two types of counterparts to BNS mergers: kilonovae and SGRBs.  Kilonovae are optical/infrared transients, lasting a few days to a \cg{few weeks}, that are powered by the radioactive decay of $r$-process nuclei synthesized in the merger ejecta \cg{(e.g.,\citealt{Metzger+10,2019MNRAS.486..672A,2020MNRAS.493.3379R})}.  The kilonova signal is expected to be relatively isotropic (however, see e.g., \citealt{Kasen+15,Fontes+20,Darbha&Kasen20}).  For simplicity, in our estimates of the EM detection horizon below, we assume that all kilonovae exhibit light curves identical to GW170817 (e.g., \citealt{Cowperthwaite+17,Villar+17a}), independent of the binary inclination.  In actuality, some diversity in the kilonova properties is expected (e.g., \citealt{Margalit&Metzger17}) and hinted at by observations (e.g., \citealt{Gompertz+18,Fong+20}).  However, we leave a detailed study of variations in the kilonova properties to future work\footnote{Once the mapping between GW and EM properties is better understood, one can envision an observing program that selects particular GW events for follow-up based on the expected kilonova properties (e.g., \citealt{Margalit&Metzger19}).}.

VRO is the most promising facility for kilonova follow-up in the 2020s. \cg{Survey operations are expected to begin by 2022-2023 with the main science survey lasting at least ten years \citep{Ivezic+19}. }    For this work, we consider a target-of-opportunity follow-up program similar to those described in \citet{Margutti+18} and \citet{Cowperthwaite+19}.  In particular, we consider a program using only two filters to minimize the observing time requirement while still providing color information for efficient kilonova identification \citep{Cowperthwaite&Berger15,Cowperthwaite+18}.

\begin{figure}
	\centering
	\includegraphics[width=1.0\columnwidth]{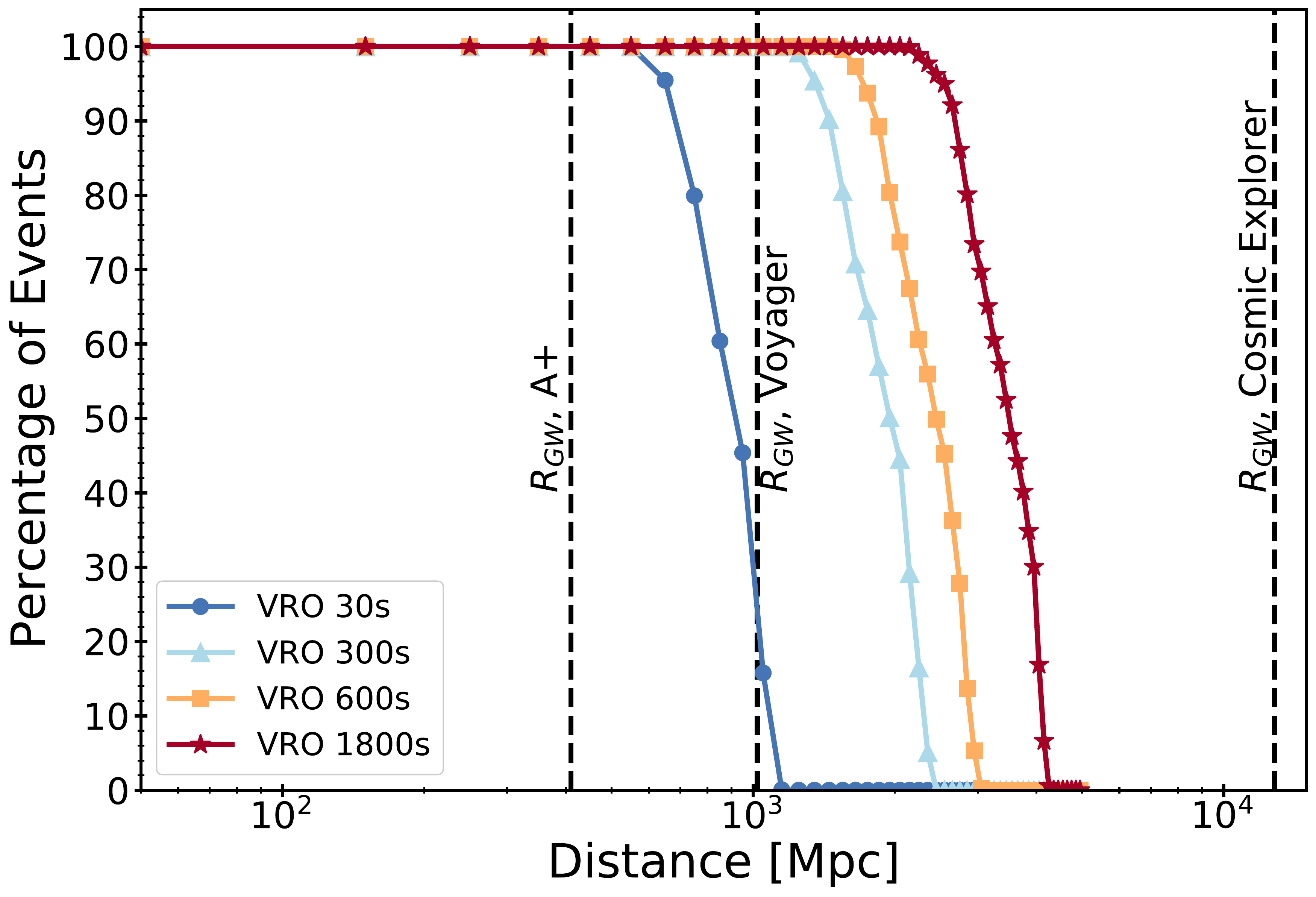}
	\caption{Kilonova detection efficiency with VRO for \cg{``VRO 30s''} (30 s exposure), \cg{``VRO 300s''} (300 s), \cg{``VRO 600s''} (600 s), and \cg{``VRO 1800s''} (1800 s) observing scenarios.  Vertical lines show the SFR Reach 50 detection range of A+ and Voyager; namely, the distance interior to which 50\% of detections occur for a BNS source population following the cosmic star formation history \citep{Chen+17}.}
	\label{fig:efficiency}
\end{figure}

We define several distinct observing scenarios for the allocation of VRO time, based on the assumed integration time per pointing (Table~\ref{tab:scenarios}).  These range from \cg{``VRO 30s''} scenarios requiring modest time allocations, to ``\cg{VRO 1800s}'' scenarios that in the Voyager/CE era (i.e., following the completion of the decade-long VRO) would take up a considerable fraction of the entire VRO science time.  Our \cg{VRO 30s} scenario adopts 30 s exposures, similar to the VRO main science survey.  Our ``\cg{VRO 300s}'', ``\cg{VRO 600s}'', and \cg{VRO 1800s} scenarios assume 300, 600, and 1800 s exposures, respectively.  Simulated observations for these scenarios were conducted using the procedure outlined in \citet{Cowperthwaite+19}.  In Figure~\ref{fig:efficiency} we show the resulting kilonova detection efficiency as a function of distance. With the detection efficiency for each scenarios we define a limiting EM observing distance, $D_{L,\rm lim}$, as the luminosity distance out to which the detection efficiency is larger than 99\%.  Also shown for comparison are BNS merger detection ranges of various GW observatories (defined as the distance inside which half of GW detections take place assuming a source rate following the global star formation history; \citealt{Chen+17}). 

Our kilonova follow-up scenarios assume three epochs of VRO observations in two filters.  We focus on events with sky localizations of $\lesssim 20$ deg${}^2$, to minimize the number of required telescope pointings to four.  We estimate the fraction of such GW events through simulations following \citet{loc2d,loc3d} and find $f_{\rm <20deg}\approx 0.6-0.95$ for different GW instruments and cuts on source distance (Table~\ref{tab:scenarios})~\footnote{\cg{We note that having KAGRA and LIGO-India in the network will change the fraction a bit.}}.  We also assume 2 min of overhead for filter changes and negligible overhead for slew time and CCD readout. Finally, taking into account sky accessibility and weather and observing conditions, we estimate that only a fraction of $f_{\rm obs}\approx 0.4$ of the detectable BNS merger population will have their counterparts and redshifts determined \citep{Cowperthwaite+18}. 

\cg{We note that a promising facility for space-based observations of kilonovae is the Nancy Grace Roman Space Telescope (Roman Space Telescope, formerly WFIRST; \citealt{Dore+19}), which is expected to launch in the late 2020's with a planned five year mission. In particular, the Roman Space Telescope will facilitate near-infrared (NIR) observations of kilonovae, probing regions of the spectral energy distribution (SED) that are potentially difficult to observe from ground-based facilities (see e.g., \citealt{Foley+19}). This capability is also crucial for obtaining observations of distant or strongly reddened kilonovae that are otherwise undetectable by telescopes like the VRO.}

In contrast to kilonovae, SGRBs and their associated afterglows are much more luminous, but the emission is relativistically beamed into a narrow solid angle fraction $f_b = 1-{\rm cos}\theta_j$, such that BNS mergers are detectable only when the binary inclination angle, $\iota$, is smaller than the jet half-opening angle, $\theta_j$; namely $\iota \lesssim \theta_j$ or $\iota \gtrsim \pi-\theta_j$. 
\cg{Here we assume the directions of the jets are aligned with the binary orbital angular momentum.}
Motivated by observations of SGRBs, we assume all BNS mergers produce GRB jets with\footnote{Although GW170817 was detected via its gamma-ray emission \citep{Goldstein+17,Savchenko+17} at an inclination angle of $2\theta_j\approx 20-25^{\circ}$ (e.g., \citealt{Margutti+17,Alexander+17}), its gamma-ray luminosity was so suppressed that the same event would not be detectable at the much larger distances of GW-detected mergers in the A+ era and beyond, even with next-generation gamma-ray telescopes.} $\theta_j = 10^{\circ}$ \citep{Berger14}.  Given our assumed local BNS rate of $\mathcal{R}(z = 0) = 300$ Gpc$^{-3}$ yr$^{-1}$ (\S\ref{sec:simulations}), the predicted local SGRB rate is thus $f_b\mathcal{R}(z = 0)\approx 5$ Gpc$^{-3}$ yr$^{-1}$, broadly consistent with SGRB observations \citep{Wanderman&Piran15}.  While relativistic beaming reduces the number of joint SGRB-GW events, these events are of a higher quality for standard siren cosmology than joint kilonova-GW events because the inclination angle of the binary will be tightly constrained ($\sigma_{\iota}\approx \theta_j\approx 10^{\circ}$) thereby reducing the uncertainty of the luminosity distance from the GW data \citep{2019PhRvX...9c1028C}.   

To date, most SGRB redshifts have been obtained for events localized with the {\it Swift} XRT \citep{Berger14}.  Our baseline SGRB scenarios therefore consider a gamma-ray satellite with capabilities similar to that of {\it Swift}.  This satellite could represent {\it Swift} itself while still operational, or a mission with comparable capabilities that flies over the next decade or longer (e.g., the Chinese-French satellite {\it SVOM}; \citealt{Cordier+15}).  Based on the current success rate of redshift determination with {\it Swift}, we assume that redshifts will be obtained for a fraction $f_z = 0.3$ of detected SGRBs.  Furthermore, the field-of-view (FOV) of {\it Swift}/BAT is 1.4 steradians, corresponding to an all-sky fraction $f_{\rm sky} = 0.11$.  Thus, even assuming every BNS merger produces an SGRB for observers within the beaming cone, only a fraction $f_{\rm obs} = f_{\rm z}\times f_{\rm sky}\approx 0.03$ of those events will have their redshifts determined.  The detection range of a {\it Swift}-like satellite ($D_{L,{\rm lim}}$ in Table~\ref{tab:scenarios}) is estimated as the distance out to which the total SGRB rate equals the detection rate of {\it Swift} of about $10$ SGRBs per year (e.g., \citealt{Burns+16}).

Several next generation gamma-ray satellites are under consideration which could improve upon the capabilities of {\it Swift} (e.g., \citealt{Camp19,McEnery+19}).  Thus, in the era of Voyager/CE we consider scenarios employing a {\it Swift}-like mission but with improved capabilities.  We define ``{\it Swift}+'' as a gamma-ray/X-ray satellite with similar sensitivity and localization capability to {\it Swift}, but with a larger FOV covering 50\% of the sky (leading to $f_{\rm obs} = 0.15$); i.e., similar to the {\it Fermi} GBM FOV but with XRT localization capability.  

Even if future gamma-ray satellites do not possess a rapid-slewing X-ray telescope similar to {\it Swift}/XRT, a redshift determination could in principle be enabled by detection of the optical afterglow either on-board or from ground-based follow-up (e.g., with VRO; see \S\ref{sec:program}).  We further define a more ambitious future mission ``{\it Swift}++'', which may overlap with CE in the 2040s, that has both the same large FOV of {\it Swift}+ and greater sensitivity, thus increasing the SGRB detection rate by a factor of 6 (similar to the capabilities of the proposed {\it AMEGO} satellite; \citealt{McEnery+19}. \cg{We note that the proposed {\it THESEUS} satellite with a different design will also lead to a factor of a few more SGRB detections; \citealt{Amati:2017npy}}).

\begin{table*}
\caption{\label{tab:scenarios}Joint GW-EM Observing Scenarios\protect \\}
\begin{centering}
\begin{tabular}{|c|c|c|c|c|c|c|}
\hline 
Counterpart & GW ($R_{\rm GW}^{(a)}$)  & VRO int. time/gamma-ray telescope ($D_{L,\rm lim}^{(b)}$) & $f_{\rm obs}^{(c)}$ & $f_{{\rm 20deg}^2}^{(d)}$   & $\dot{N}_{\rm GW/EM}^{(e)}$ & $\mathcal{F}_{\rm obs}^{(f)}$\tabularnewline
\hline
KN & A+ (410 Mpc)  &  VRO 30 s (575 Mpc)  & 0.4 & 0.8  & 12 & 0.0008\tabularnewline
KN & Voyager (1020 Mpc) & VRO 30 s (575 Mpc)  & 0.4 & 0.8 & 28 & 0.002\tabularnewline
KN & Voyager (1020 Mpc) & VRO 300 s (1250 Mpc)  & 0.4 & 0.7 & 114 & 0.06\tabularnewline
KN & Voyager (1020 Mpc) & VRO 1800 s (2250 Mpc)  & 0.4 & 0.6 & 144 & 0.48\tabularnewline
KN & CE (1.284 Gpc) & VRO 30 s (575 Mpc) & 0.4 & 1.0  & 39 & 0.003\tabularnewline
KN & CE (1.284 Gpc) & VRO 300 s (1250 Mpc)  & 0.4 & 0.95  & 321 &  0.18 \tabularnewline
KN & CE (1.284 Gpc) & VRO 600 s (1550 Mpc)  & 0.4  & 0.95 & 572 &  0.6 \tabularnewline
KN & CE (1.284 Gpc) & VRO(+) 1800 s (2250 Mpc)  & 0.4   & 0.9  & 300(1425) &  1(4.75) \tabularnewline
\hline
GRB & A+ (410 Mpc) & Swift (3 Gpc)  & 0.03 & N/A  & 0.07 &  $ \ll 1$\tabularnewline
GRB & A+ (410 Mpc) & Swift+ (3 Gpc) & 0.15 &  N/A & 0.35 &  $ \ll 1$\tabularnewline
GRB & Voyager (1020 Mpc) & Swift (3 Gpc) & 0.03 & N/A & 1 & $ \ll 1$\tabularnewline
GRB & Voyager (1020 Mpc) & Swift+ (3 Gpc)  & 0.15 & N/A  & 5& $ \ll 1$ \tabularnewline
GRB & CE (1.284 Gpc)  & Swift (3 Gpc) & 0.03 & N/A & 3 & $ \ll 1$\tabularnewline
GRB & CE (1.284 Gpc)  & Swift+ (3 Gpc) & 0.15 & N/A & 16 & $ \ll 1$\tabularnewline
GRB & CE (1.284 Gpc) & Swift++ (5.6 Gpc) & 0.15 & N/A & 91 & $ \ll 1$\tabularnewline
\hline
\end{tabular}
\par\end{centering}
$^{(a)}$Distance within which half of GW sources are detected (SFR Reach 50; see definition in \citealt{Chen+17}).
\\$^{(b)}$For KN, distance out to which the detection efficiency is larger than 99\% (Fig.~\ref{fig:efficiency}). For GRB, distance out to which 
the all-sky GRB rate equals $\dot{N}_{\rm GW/EM}^{(h)}f_{\rm obs}^{-1}$. 
\\$^{(c)}$Efficiency of identifying EM counterpart and redshift for events in the joint EM/GW sensitivity volume.  In the case of Rubin Observatory this accounts for e.g. bad weather or an inaccessible sky position.  In the GRB case it accounts for the limited field of view of the gamma-ray detector and inefficiencies in obtaining a redshift from the afterglow (but {\it not} for the jet beaming fraction). 
\\$^{(d)}$Fraction of GW sources within $D_{L,\rm lim}$ that are localized to better than 20 deg$^2$.  
\\$^{(e)}$Number of joint GW/EM detections per year.
\\$^{(f)}$Fraction of telescope time dedicated to GW/EM follow-up program.  We have assumed 3600 hours total time per year available to the Rubin Observatory and 7900 hours available to GRB telescopes (>90\% duty cycle for orbit similar to {\it Swift}).
\end{table*}

\begin{figure}
	\centering
	\includegraphics[width=1.0\columnwidth]{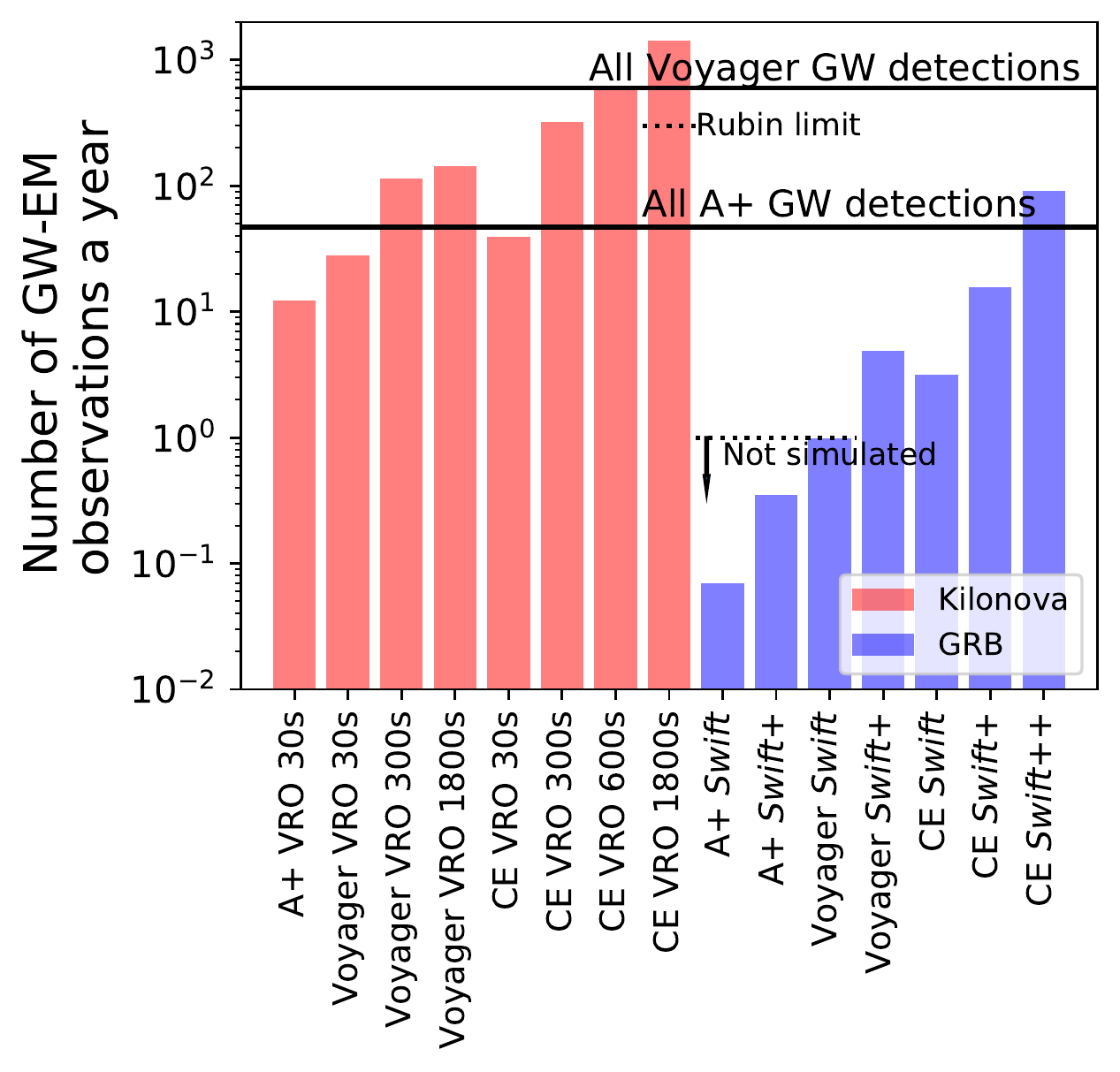}
	\caption{Yearly rate of joint GW-EM detections for the various scenarios in Table~\ref{tab:scenarios}. For comparison we also show the total number of GW detections by A+ and Voyager.  Due to their low joint detection rates of $\lesssim 1$ yr$^{-1}$, we do not include the first three SGRB scenarios in our standard siren analysis.  We also cap the number of joint detections in our \emph{CE \cg{VRO 1800s}} scenario as marked, so as not to require more than 100\% of the available VRO time.}
	\label{fig:ndetection}
\end{figure}

\section{Simulations}
\label{sec:simulations}

We consider a population of BNS mergers with a local volumetric rate of $\mathcal{R}(z = 0) = 300$ Gpc$^{-3}$ yr$^{-1}$, close to the median of the observed rate by LIGO-Virgo~\citep{Abbott:2020gyp} and in line with the beaming-corrected SGRB rate (e.g., \citealt{Fong+17}). We assume that the BNS rate tracks the global star formation at $z\gtrsim 0$ \citep{Madau&Dickinson14}, as supported by the host galaxy properties of SGRBs \citep{Berger14}.  Under these assumptions, the predicted total detection rates of BNS mergers by A+, Voyager, and ET are about 50, 600, and $3.5\times 10^4$ yr$^{-1}$, respectively.  

For each scenario considered in Table~\ref{tab:scenarios} we begin with $10^4$ $1.4\msun-1.4\msun$ non-spinning BNS detection simulations using the \imr waveform\cg{~\citep{2016PhRvD..93d4006H,2016PhRvD..93d4007K}}. A GW detection threshold is set at network signal-to-noise of 12.  We use the standard cosmology from Planck, with $H_0=0.679$, $\Omega_m=0.3065$, $\Omega_k=0$, and $w=-1$ \citep{Planck2016}, and adopt a flat \textit{w}CDM ($\Omega_k=0$) model.  We then randomly select $\dot{N}_{\rm GW/EM}$ number of events specified in Table~\ref{tab:scenarios} for the standard siren measurement described below. We repeat this random selection 20 times and report the average and standard deviation of the cosmological constraints in \S\ref{sec:results}.
 
To estimate the cosmological parameters, we follow the Bayesian framework of \citet{2020arXiv200914057C} and use \texttt{emcee} \citep{2013PASP..125..306F} to estimate the posteriors of $H_0$, $\Omega_m$, and $w_0$. For an event with GW data ($\mathcal{D}_{\rm GW}$) and EM data ($\mathcal{D}_{\rm EM}$), the posterior of $(H_0,\Omega_m,w_0)$ can be written as:
\begin{equation}
\begin{split}
\label{eq:posterior}
&p(H_0,\Omega_m,w|\mathcal{D}_{\rm GW},\mathcal{D}_{\rm EM}) = p(H_0,\Omega_m,w_0)\times \\
&\frac{\displaystyle\int p(\mathcal{D}_{\rm GW}|\vec{\Theta})p(\mathcal{D}_{\rm EM}|\vec{\Theta})p_{\rm pop}(\vec{\Theta}|H_0,\Omega_m,w_0)d\vec{\Theta}}{\displaystyle\int p_{\rm det}(\vec{\Theta})p_{\rm pop}(\vec{\Theta}|H_0,\Omega_m,w_0)d\vec{\Theta}}\;,
\end{split}
\end{equation}
where $\vec{\Theta}$ represents all the binary parameters, such as the masses, spins, luminosity distance ($D_L$), sky location, inclination angle ($\iota$); and $p(H_0,\Omega_m,w_0)$ is the prior probability density function on the cosmological parameters.  All of our parameter priors are motivated by cosmology results obtained previously through the standard siren GW technique.  We begin with a $5\%$ Gaussian $H_0$ prior centering around the simulated value, motivated by the precision expected by the A+ era \citep{Chen+17}. For $\Omega_m$ we apply a flat prior between $[0,1]$, and for $w_0$ a flat prior between $[-2,-0.3]$. \cg{The  population prior $p_{\rm pop}(\vec{\Theta}|H_0,\Omega_m,w_0)$ is the probability density of binaries with parameters $\vec{\Theta}$ under our assumption of rate evolution (i.e. tracking the global star formation; \citealt{Madau&Dickinson14}) in the Universe with parameters $(H_0,\Omega_m,w_0)$. } 

The detection probability is given by:
\begin{equation}
\label{eq:det}
p_{\rm det}(\vec{\Theta})\equiv \displaystyle \iint\limits_{\substack{{\mathcal{D}_{\rm GW}>{\rm GW}_{\rm th}}, \\{\mathcal{D}_{\rm EM}>{\rm EM}_{\rm th}}}} p(\mathcal{D}_{\rm GW}|\vec{\Theta})p(\mathcal{D}_{\rm EM}|\vec{\Theta})d\mathcal{D}_{\rm GW}d\mathcal{D}_{\rm EM}\;, 
\end{equation}  
in which the integration is only carried out over data above the GW and EM detection thresholds, ${\rm GW}_{\rm th}$ and ${\rm EM}_{\rm th}$, respectively. In our simulations the thresholds are determined by the GW network signal-to-noise ratio of 12, the EM observable distance limit $D_{L,\rm lim}$, and the range of binary inclination $\iota_{\rm GRB}$ (only applicable in the SGRB cases). 

We assume that when an EM counterpart is detected, the redshift and sky location of the BNS are precisely determined.  
If an SGRB is observed, we assume $\iota$ is measured with a Gaussian uncertainty $\sigma_{\iota}=10^{\circ}$. Under our assumptions, the GW likelihood $p(\mathcal{D}_{\rm GW}|\vec{\Theta})$ in Equation~\ref{eq:posterior} is reduced to the GW distance-inclination angle likelihood $p(\mathcal{D}_{\rm GW}|D_L,\iota)$ fixed along the BNS's sky location. Therefore we can use the rapid GW distance-inclination angle estimation algorithm developed in \cite{2019PhRvX...9c1028C} to simulate the likelihood. 

Figure~\ref{fig:ndetection} and Table~\ref{tab:scenarios} show that the rate of joint GW-EM detections is substantially smaller than the total GW detection rate due to limitations imposed by EM capabilities. This is particularly acute for CE, which can detect BNS mergers to substantial redshifts.  In particular, there are more events than VRO can reasonably follow up in the \emph{CE \cg{VRO 1800s}} scenario.  We scale down the number of events to 300 per year assuming mature observing strategies will be formulated by then and help in down-selecting the most impactful mergers.  On the other hand, less than one joint GW-GRB detection per year is expected for the \emph{A+ \cg{\it{Swift}}}, \emph{A+ \cg{\it{Swift}+}} and \emph{Voyager \cg{\it{Swift}}} scenarios, so we eliminate these in the standard siren simulations.  

\section{Results and Discussion}
\label{sec:results}

\begin{figure}
	\centering
	\includegraphics[width=1.0\columnwidth]{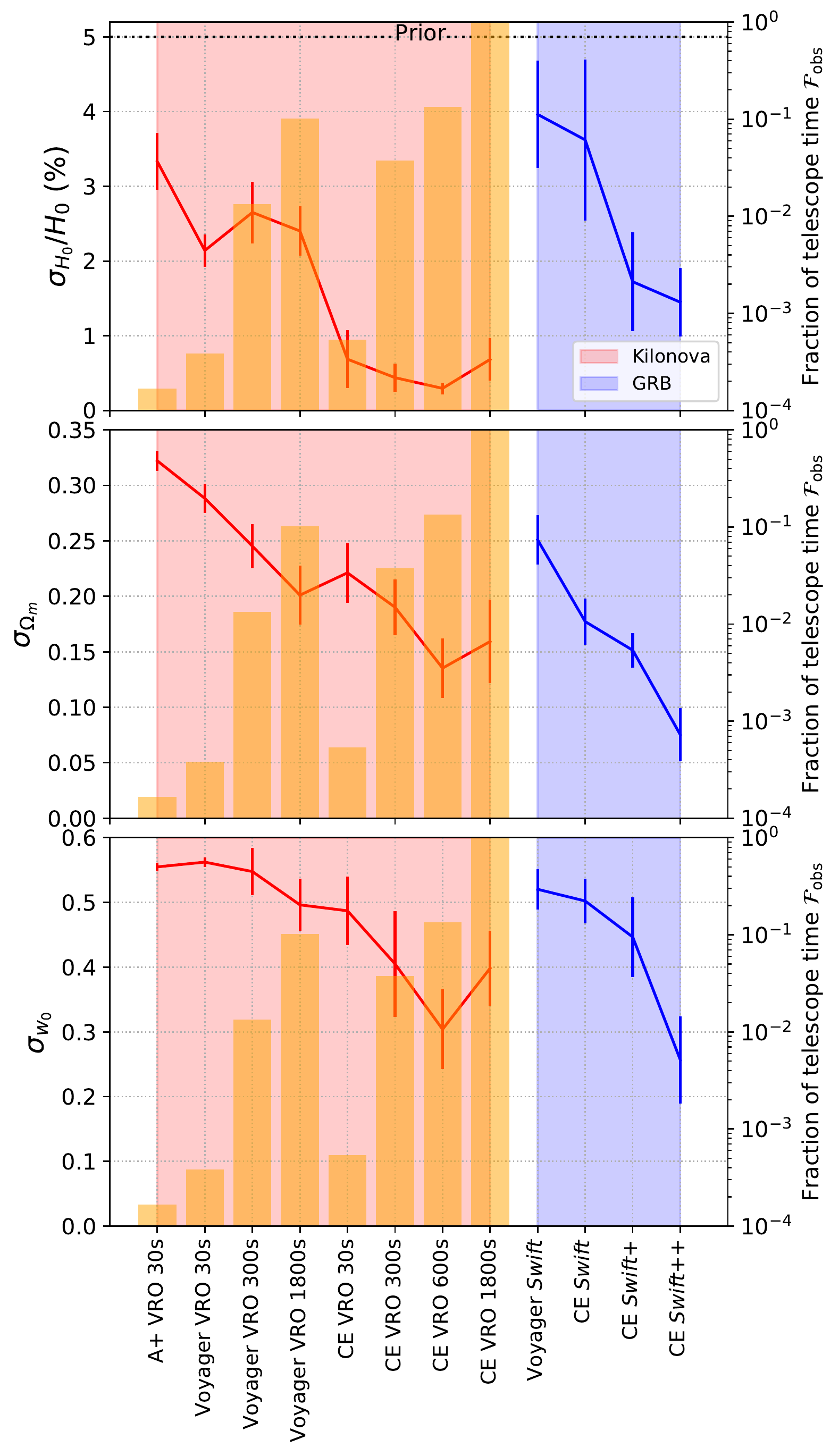}
	\caption{The symmetric 68\% uncertainty of $(H_0,\Omega_m,w_0)$ for each scenario in Table~\ref{tab:scenarios} ($H_0$ is reported in fractional uncertainty for comparison to literature). The error bars show the standard deviation of the uncertainties over 20 repeated simulations.  The orange bars (right vertical axis) indicate the fraction of the total observing time available to the Rubin Observatory for each kilonova scenario.  }
	\label{fig:summary}
\end{figure}

Our key results are summarized in Figure~\ref{fig:summary}, which shows the symmetric 68\% uncertainties in the parameters $(H_0,\Omega_m,w_0)$ achieved for each of our scenarios, assuming one year of joint GW-EM observations.  We focus on one parameter at a time and marginalize over the other two in the posteriors (Equation~\ref{eq:posterior}).  For the kilonova scenarios, we also show the required fraction of VRO time for reference with orange bars.\footnote{This is not relevant for the SGRB scenarios since we assume that the SGRB detection will occur as a part of routine operations.} Our results for the kilonova scenarios can be summarized as follows:
\begin{itemize}
    \item \emph{A+} and \emph{Voyager \cg{VRO 30s}}: The events captured in these scenarios are nearby, so they provide a constraint on $H_0$ but not on $\Omega_m$ and $w_0$. On the other hand, they require only a small fraction, $\lesssim 1\%$ of VRO observing time.
    
    \item \emph{Voyager \cg{VRO 300s}} and \emph{\cg{VRO 1800s}}: The larger distances of the joint GW-EM detections enabled by the more substantial investment of VRO time provide appreciable constraints on $(\Omega_m,w_0)$.  The number of joint detections and the results from both scenarios are comparable.  However, \emph{\cg{VRO 1800s}} requires eight times more VRO time than \emph{\cg{VRO 300s}}. The \emph{\cg{VRO 1800s}} scenario is essentially ``overkill'' because the KN sensitivity distance greatly exceeds the GW one (Figure~\ref{fig:efficiency}) and hence the rate of joint detections is only marginally higher in the \emph{\cg{VRO 1800s}} than in the \emph{\cg{VRO 300s}} case (Fig.~\ref{fig:ndetection}). 
    
    \item \emph{CE \cg{VRO 30s}}: Even with fewer joint events than \emph{Voyager \cg{VRO 300s}}/\emph{\cg{VRO 1800s}}, this scenario constrains $H_0$ to a sub-percent level, because CE measures the source luminosity distances to much greater precision than A+/Voyager.  Given the small number of events, this scenario requires $\lesssim 1\%$ of VRO time.  On the other hand, the constraints on $(\Omega_m,w_0)$ are comparable to the \emph{Voyager \cg{VRO 300s}} and \emph{\cg{VRO 1800s}} scenarios because \emph{CE KN \cg{VRO 30s}} has a limited horizon of about 575 Mpc.
    
    \item \emph{CE \cg{VRO 300s}} and \emph{\cg{VRO 600s}}: These scenarios lead to better constraints on $(\Omega_m,w_0)$ compared to the \cg{\emph{VRO 30s}} scenario, but the telescope time needed also become 60 and 200 times higher than the \cg{\emph{VRO 30s}} scenario.
    
    \item \emph{CE \cg{VRO 1800s}} scenario: After scaling the number of events down to 300 so as not to exceed 100\% of VRO time, this scenario actually provides worse constraints than the \emph{\cg{VRO 600s}} scenarios due to a reduction in the number of events. The \emph{\cg{VRO 1800s}} scenario is overkill since the events at 2 Gpc do not provide more constraints than events within 1.5 Gpc observed in the \emph{\cg{VRO 600s}} scenario.  
\end{itemize}

Our results for the SGRB scenarios (with $\gtrsim 1$ joint detection per year) can be summarized as follows:
\begin{itemize}
    \item \emph{Voyager \cg{\it{Swift}+}}:  Even with only a few GRBs, the constraint obtained on $\Omega_m$ is comparable to the \emph{Voyager \cg{VRO 300s}} and \emph{\cg{VRO 1800s}} scenarios.  This is mainly because the luminosity distance uncertainty is reduced due to constraints on the binary inclination from the SGRB detection.  However, due to the lack of nearby events given the lower number of events, the constraint on $H_0$ only marginally improves over the prior.
    
    \item \emph{CE \cg{\it{Swift}}} and \emph{\cg{\it{Swift}+}}: Similarly, only a few SGRB events are sufficient to achieve comparable precision on $(\Omega_m,w_0)$ measurement as the \emph{CE \cg{VRO 600s}} or \emph{\cg{VRO 1800s}} scenarios. In addition to the benefit of the inclination angle constraint, the SGRB sample reaches to higher redshifts than kilonovae in the CE era.
    
    \item \emph{CE \cg{\it{Swift}++}}: Given the greater number of SGRBs and their larger detection distance, $(\Omega_m,w_0)$ can be measured to $20-30\%$ precision over one year of observations.  Unlike in the kilonova case, even in the \cg{\it{Swift}++}  scenario, we are not limited by the gamma-ray telescope time (though constraints on spectroscopic follow-up may be more severe; see below).
\end{itemize}

Although we report results for only one year of observations, a longer observing period is of course likely. For example, we find that with five years of observation, uncertainties on the cosmological parameters improve by a factor of about $1.5-2$.
 
\subsection{Uncertainties and Caveats}

Motivated empirically by observations with {\it Swift}, we have assumed that $30\%$ of SGRBs detected by the gamma-ray satellite will have their redshift precisely determined.  While it is potentially realistic to obtain spectroscopic follow-up of SGRB afterglows or host galaxies in the era of a few (or even a few dozen) events per year, our \emph{CE \cg{\it{Swift}++}} scenario would require about $100$ redshift measurements per year, possibly from 8-m class or larger spectroscopic telescopes (given the large distances of SGRBs) by the 2040s when CE is operational.  
If photometric redshifts are required for large samples, this can introduce additional systematic uncertainties not accounted for in our calculations. Likewise, our more ambitious \emph{Voyager/CE VRO 1800s} scenarios require obtaining hundreds of host galaxy redshifts, albeit for more nearby events, which might enable the use of smaller (2-m and 4-m class) telescopes.

Although we consider each of the scenarios above as taking place independently, a string of successively-executed scenarios (\cg{e.g., if the {\it A+ VRO 30s} scenario is followed by the {\it Voyager 1800s} scenario, the cosmological measurement posterior at the end of the {\it A+ VRO 30s} program will serve as the prior of the {\it Voyager 1800s} program.}) will also lead to tighter constraints.  Furthermore, with our only assumed priors being that on $H_0$ from an assumed Advanced LIGO-Virgo standard siren program, our results correspond to an independent measurement of cosmological parameters made using just the standard siren method.  However, assuming that systematic uncertainties are under control, other GW-only measurements (e.g., \citealt{2019ApJ...883L..42F}) or non-GW measurement (e.g., CMB) could potentially provide greater leverage at high redshift and be combined to better measure cosmological parameters. It has been shown that a synergy between the standard siren and CMB+BAO+Type Ia supernovae will lead to a factor of $\sim 2$ tighter constraints \citep{2020JCAP...03..051J}. 

Known possible systematic uncertainties for the standard siren method, including the instrumental calibration uncertainty \citep{2020arXiv200502531S}, the GW and EM selection effect \citep{2019MNRAS.486.1086M,2020arXiv200602779C}, the use of photometric redshift mentioned above, and the lensing of GW signals, are not included in our analysis. These systematics can introduce percent level uncertainty on the $H_0$ measurement, competing with the statistical ones.  As GW instruments evolve and our understanding of the BNS population and their EM counterparts improves, it is reasonable to believe that these systematics will become better controlled.  

For simplicity we have considered a sharp cutoff in the detection horizon $D_{L,\rm lim}$ of kilonovae and SGRBs, but the actual detection efficiency will taper with source distance more gradually (e.g., Figure~\ref{fig:efficiency}).  Likewise, in calculating required VRO time, we have assumed a uniform per-source integration time for all kilonova searches independent of distance.  In reality, nearby events, or those with particularly luminous kilonovae, may require fewer resources.  A more realistic EM detection efficiency can be constructed after a concrete observing plan is in place.  Overall, the systematic effect on $(\Omega_m,w_0)$ measurements are less significant than for $H_0$ due to their larger statistical uncertainties.

\section{Recommended Program}
\label{sec:program}

\begin{table}
	\centering
	\begin{tabular}{c c  c  c c}
		 Scenario & $\sigma_{H_0}/H_0$ (\%)   & $\sigma_{\Omega_m}$  & $\sigma_w$   \\
		\hline
		 \cg{A+ VRO 30s} & 3.3(2.2) &0.32(0.21) &0.55(0.37) \\ 
		 \cg{Voyager VRO 300s} & 2.7(1.8) &0.25(0.16) &0.55(0.37) \\ 
		 \cg{CE VRO 600s} &0.3(0.2) &0.14(0.09) &0.30(0.20)   \\ 
		 \cg{Voyager \it{Swift}+}&4.0(2.6) &0.25(0.17) &0.52(0.35) \\ 
	     \cg{CE \it{Swift}++} & 1.4(1.0) &0.08(0.05) &0.26(0.17)    
	\end{tabular}
	\addtolength{\tabcolsep}{-10pt}
	\caption{Joint GW and kilonova/SGRB observing programs selected from Table~\ref{tab:scenarios} and their constraints on cosmological parameters over 1 year (5 year) of observations. The 5-year constraints are conservatively taken to be a factor of 1.5 better than our simulated 1 year observations. \label{tab:recommend}}
\end{table}

In summary, for kilonovae programs, improved $H_0$ precision is largely driven by upgrades in the GW facility rather than a more ambitious EM follow-up program.  This is because the greatest leverage on $H_0$ comes from the nearest GW events with the highest SNR, for which VRO can most easily detect the kilonova.  Increasing the sample to events at greater distances (\emph{\cg{VRO 300s}} and \emph{\cg{VRO 1800s}} programs) requires significantly greater observing time VRO, for only a marginal improvement in constraints.  On the other hand, once in the CE era, collecting a large sample of kilonovae at \emph{\cg{VRO 600s}} redshifts does provide noticeable improvements on $(\Omega_m, w_0$).

Due to the required on-axis orientation, the number of joint GW-SGRB observations does not become appreciable until the Voyager/CE eras.  However, with even a few events, SGRBs can lead to comparable or better constraints than the kilonova scenarios for $(\Omega_m, w_0$).  By contrast, none of the SGRB scenarios compete with the kilonova scenarios in measuring $H_0$ for the same GW detector configuration.  The constraints on $(\Omega_m, w_0$) largely scale with the total number of GW/SGRB joint detections.  Thus it is no surprise that a future gamma-ray satellite with both a greater sensitivity and larger field of view than {\it Swift} (i.e., {\it `` Swift++''}) performs better than one with just a larger field of view ({\it ``Swift+''}).

In light of these findings, we propose the following programmatic guidelines.  
\begin{itemize}
    \item VRO should undertake an active kilonova target-of-opportunity follow-up program (in A+ and beyond) focused on the nearest GW events (outside the distance afflicted by peculiar velocity uncertainties), as these provide the greatest leverage on $H_0$.  For purposes of cosmological studies alone, it is not necessarily of benefit to target the greater number of events in the \emph{Voyager/CE} eras at greater distances (and hence dimmer kilonovae) because of the diminishing returns per invested telescope time.  
    On the other hand, observations of kilonovae may be motivated by other, non-cosmology-related science cases, such as $r$-process nucleosynthesis or host galaxy properties.  Kilonovae programs targeting sources at larger distance can provide $(\Omega_m, w_0$) constraints, particularly absent the gamma-ray facility recommended below. 
    
    \item To fully extend the standard siren technique to $(\Omega_m, w_0$) studies, it is important to have an active gamma-ray satellite with {\it Swift}-like GRB localization capabilities to coincide with Voyager and CE in the 2030s and beyond.  Ideally this future instrument would have both larger FOV and greater sensitivity than {\it Swift}/BAT, since the returns scale with the total number of SGRBs with redshifts.  Concepts for future gamma-ray satellites with greater sensitivity (e.g., AMEGO; \citealt{McEnery+19}) and/or wider FOV/on-board localization capabilities (e.g., TAP; \citealt{TAP+19}) than {\it Swift} have recently been proposed to NASA as future missions.  
    
    If on-board localization capabilities are not feasible, an alternative approach is to localize and identify the merger host galaxy/redshift by detecting the SGRB optical afterglow from the ground (e.g., with VRO), particularly if the angular resolution of gamma-ray instrument is less than a few degrees.  However, exploring this possibility in depth is beyond the scope of this paper; such an approach could end up requiring as much or more observing time as the proposed KN programs.  
    
    \item In parallel, a plan should be developed to increase the capacity to obtain host galaxy redshifts for GW/EM events. In the GRB joint-detection case, this represents an expansion from  $\sim$ few events per year, to tens or even hundreds per year detected by next-generation facilities.  Likewise, for kilonovae, several hundreds of events will be required per year by the CE/ET era. One possibility is the expansion of an ongoing spectroscopic survey such as the Dark Energy Spectroscopic Instrument (DESI). Additionally, photometric redshifts measured by upcoming survey telescopes like VRO are another avenue for obtaining a large sample of redshifts.  However, additional work (also required for other areas of cosmology) is necessary to reduce the statistical and systematic uncertainties in photometric redshifts to the sub-percent level expected for $H_0$ constraints by the CE era.
    \end{itemize}
    
Our recommended programs, and their projected constraints on cosmological parameters with one(five) years of observations, are summarized in Table~\ref{tab:recommend}.

In conclusion, this work has explored the prospects of the GW-EM standard siren method to address key questions in cosmology over the next three decades.  With future upgrades of the GW and EM facilities, and coordinated efforts between these communities starting now, multi-messenger standard siren cosmology has great potential to provide insights into the expansion history of the Universe.

\acknowledgements{
We acknowledge valuable discussions with Matthew Evans. H.-Y.~C. was supported by the Black Hole Initiative at Harvard University, through a grant from the John Templeton Foundation and the Gordon and Betty Moore Foundation. H.-Y.~C. and P.~S.~C. are grateful for support provided by NASA through the NASA Hubble Fellowship grant \#HST-HF2-51452.001-A and \#HST-HF2-51404.001-A awarded by the Space Telescope Science Institute, which is operated by the Association of Universities for Research in Astronomy, Inc., for NASA, under contract NAS 5-26555.  B.~D.~M. is supported in part by NSF grant AST-2002577 and NASA grant NNX17AK43G.  }

\bibliographystyle{yahapj}
\bibliography{references}

\begin{thebibliography}{}
\providecommand\natexlab[1]{#1}
\providecommand\JournalTitle[1]{#1}

\bibitem[{{Abazajian} {et~al.}(2019){Abazajian}, {Addison}, {Adshead}, {Ahmed},
  {Allen}, {Alonso}, {Alvarez}, {Anderson}, {Arnold}, {Baccigalupi}, {Bailey},
  {Barkats}, {Barron}, {Barry}, {Bartlett}, {Basu Thakur}, {Battaglia},
  {Baxter}, {Bean}, {Bebek}, {Bender}, {Benson}, {Berger}, {Bhimani},
  {Bischoff}, {Bleem}, {Bocquet}, {Boddy}, {Bonato}, {Bond}, {Borrill},
  {Bouchet}, {Brown}, {Bryan}, {Burkhart}, {Buza}, {Byrum}, {Calabrese},
  {Calafut}, {Caldwell}, {Carlstrom}, {Carron}, {Cecil}, {Challinor}, {Chang},
  {Chinone}, {Cho}, {Cooray}, {Crawford}, {Crites}, {Cukierman}, {Cyr-Racine},
  {de Haan}, {de Zotti}, {Delabrouille}, {Demarteau}, {Devlin}, {Di Valentino},
  {Dobbs}, {Duff}, {Duivenvoorden}, {Dvorkin}, {Edwards}, {Eimer}, {Errard},
  {Essinger-Hileman}, {Fabbian}, {Feng}, {Ferraro}, {Filippini}, {Flauger},
  {Flaugher}, {Fraisse}, {Frolov}, {Galitzki}, {Galli}, {Ganga}, {Gerbino},
  {Gilchriese}, {Gluscevic}, {Green}, {Grin}, {Grohs}, {Gualtieri}, {Guarino},
  {Gudmundsson}, {Habib}, {Haller}, {Halpern}, {Halverson}, {Hanany},
  {Harrington}, {Hasegawa}, {Hasselfield}, {Hazumi}, {Heitmann}, {Henderson},
  {Henning}, {Hill}, {Hlozek}, {Holder}, {Holzapfel}, {Hubmayr},
  {Huffenberger}, {Huffer}, {Hui}, {Irwin}, {Johnson}, {Johnstone}, {Jones},
  {Karkare}, {Katayama}, {Kerby}, {Kernovsky}, {Keskitalo}, {Kisner}, {Knox},
  {Kosowsky}, {Kovac}, {Kovetz}, {Kuhlmann}, {Kuo}, {Kurita}, {Kusaka},
  {Lahteenmaki}, {Lawrence}, {Lee}, {Lewis}, {Li}, {Linder}, {Loverde},
  {Lowitz}, {Madhavacheril}, {Mantz}, {Matsuda}, {Mauskopf}, {McMahon},
  {McQuinn}, {Meerburg}, {Melin}, {Meyers}, {Millea}, {Mohr}, {Moncelsi},
  {Mroczkowski}, {Mukherjee}, {M{\"u}nchmeyer}, {Nagai}, {Nagy}, {Namikawa},
  {Nati}, {Natoli}, {Negrello}, {Newburgh}, {Niemack}, {Nishino}, {Nordby},
  {Novosad}, {O'Connor}, {Obied}, {Padin}, {Pandey}, {Partridge}, {Pierpaoli},
  {Pogosian}, {Pryke}, {Puglisi}, {Racine}, {Raghunathan}, {Rahlin},
  {Rajagopalan}, {Raveri}, {Reichanadter}, {Reichardt}, {Remazeilles}, {Rocha},
  {Roe}, {Roy}, {Ruhl}, {Salatino}, {Saliwanchik}, {Schaan}, {Schillaci},
  {Schmittfull}, {Scott}, {Sehgal}, {Shandera}, {Sheehy}, {Sherwin},
  {Shirokoff}, {Simon}, {Slosar}, {Somerville}, {Spergel}, {Staggs}, {Stark},
  {Stompor}, {Story}, {Stoughton}, {Suzuki}, {Tajima}, {Teply}, {Thompson},
  {Timbie}, {Tomasi}, {Treu}, {Tristram}, {Tucker}, {Umilt{\`a}}, {van
  Engelen}, {Vieira}, {Vieregg}, {Vogelsberger}, {Wang}, {Watson}, {White},
  {Whitehorn}, {Wollack}, {Kimmy Wu}, {Xu}, {Yasini}, {Yeck}, {Yoon}, {Young},
  \& {Zonca}}]{Abazajian+19}
{Abazajian}, K., {Addison}, G., {Adshead}, P., {et~al.} 2019,
  \JournalTitle{arXiv e-prints}, arXiv:1907.04473

\bibitem[{{Abbott} {et~al.}(2016){Abbott}, {Abbott}, {Abbott}, {Abernathy},
  {Acernese}, {Ackley}, {Adams}, {Adams}, {Addesso}, {Adhikari}, \&
  et~al.}]{LIGOLocalization}
{Abbott}, B.~P., {Abbott}, R., {Abbott}, T.~D., {et~al.} 2016,
  \href{http://dx.doi.org/10.1007/lrr-2016-1}{\JournalTitle{Living Reviews in
  Relativity}, 19, 1}

\bibitem[{{Abbott} {et~al.}(2017{\natexlab{a}}){Abbott}, {Abbott}, {Abbott},
  {Acernese}, {Ackley}, {Adams}, {Adams}, {Addesso}, {Adhikari}, {Adya}, \&
  et~al.}]{LIGOH0}
---. 2017{\natexlab{a}},
  \href{http://dx.doi.org/10.1038/nature24471}{\JournalTitle{\nat}, 551, 85}

\bibitem[{{Abbott} {et~al.}(2017{\natexlab{b}}){Abbott}, {Abbott}, {Abbott},
  {Abernathy}, {Ackley}, {Adams}, {Addesso}, {Adhikari}, {Adya}, {Affeldt}, \&
  et~al.}]{LIGOG3Ref}
---. 2017{\natexlab{b}},
  \href{http://dx.doi.org/10.1088/1361-6382/aa51f4}{\JournalTitle{Classical and
  Quantum Gravity}, 34, 044001}

\bibitem[{{Abbott} {et~al.}(2017{\natexlab{c}}){Abbott}, {Abbott}, {Abbott},
  {Acernese}, {Ackley}, {Adams}, {Adams}, {Addesso}, {Adhikari}, {Adya}, \&
  et~al.}]{LIGOGW170817grb}
---. 2017{\natexlab{c}},
  \href{http://dx.doi.org/10.3847/2041-8213/aa920c}{\JournalTitle{\apjl}, 848,
  L13}

\bibitem[{{Abbott} {et~al.}(2017{\natexlab{d}}){Abbott}, {Abbott}, {Abbott},
  {Acernese}, {Ackley}, {Adams}, {Adams}, {Addesso}, {Adhikari}, {Adya}, \&
  et~al.}]{LIGOGW170817}
---. 2017{\natexlab{d}},
  \href{http://dx.doi.org/10.1103/PhysRevLett.119.161101}{\JournalTitle{Physical
  Review Letters}, 119, 161101}

\bibitem[{{Abbott} {et~al.}(2017{\natexlab{e}}){Abbott}, {Abbott}, {Abbott},
  {Acernese}, {Ackley}, {Adams}, {Adams}, {Addesso}, {Adhikari}, {Adya}, \&
  et~al.}]{LIGO+17CAPSTONE}
---. 2017{\natexlab{e}},
  \href{http://dx.doi.org/10.3847/2041-8213/aa91c9}{\JournalTitle{\apjl}, 848,
  L12}

\bibitem[{Abbott {et~al.}(2019)}]{LIGOScientific:2018mvr}
Abbott, B.~P., {et~al.} 2019,
  \href{http://dx.doi.org/10.1103/PhysRevX.9.031040}{\JournalTitle{Phys. Rev.},
  X9, 031040}

\bibitem[{Abbott {et~al.}(2020)}]{Abbott:2020gyp}
Abbott, R., {et~al.} 2020, \href{http://arxiv.org/abs/2010.14533}{{\sffamily
  arXiv:2010.14533 [astro-ph.HE]}}

\bibitem[{Ackley {et~al.}(2020)}]{Ackley:2020qkz}
Ackley, K., {et~al.} 2020,
  \href{http://dx.doi.org/10.1051/0004-6361/202037669}{\JournalTitle{Astron.
  Astrophys.}, 643, A113}

\bibitem[{{Alexander} {et~al.}(2017){Alexander}, {Berger}, {Fong}, {Williams},
  {Guidorzi}, {Margutti}, {Metzger}, {Annis}, {Blanchard}, {Brout}, {Brown},
  {Chen}, {Chornock}, {Cowperthwaite}, {Drout}, {Eftekhari}, {Frieman}, {Holz},
  {Nicholl}, {Rest}, {Sako}, {Soares-Santos}, \& {Villar}}]{Alexander+17}
{Alexander}, K.~D., {Berger}, E., {Fong}, W., {et~al.} 2017,
  \href{http://dx.doi.org/10.3847/2041-8213/aa905d}{\JournalTitle{\apjl}, 848,
  L21}

\bibitem[{Amati {et~al.}(2018)}]{Amati:2017npy}
Amati, L., {et~al.} 2018,
  \href{http://dx.doi.org/10.1016/j.asr.2018.03.010}{\JournalTitle{Adv. Space
  Res.}, 62, 191}

\bibitem[{{Andreoni} {et~al.}(2020){Andreoni}, {Goldstein}, {Kasliwal},
  {Nugent}, {Zhou}, {Newman}, {Bulla}, {Foucart}, {Hotokezaka}, {Nakar},
  {Nissanke}, {Raaijmakers}, {Bloom}, {De}, {Jencson}, {Ward}, {Ahumada},
  {Anand}, {Buckley}, {Caballero-Garc{\'\i}a}, {Castro-Tirado}, {Copperwheat},
  {Coughlin}, {Cenko}, {Gromadzki}, {Hu}, {Karambelkar}, {Perley}, {Sharma},
  {Valeev}, {Cook}, {Fremling}, {Kumar}, {Taggart}, {Bagdasaryan}, {Cooke},
  {Dahiwale}, {Dhawan}, {Dobie}, {Gatkine}, {Golkhou}, {Goobar}, {Chaves},
  {Hankins}, {Kaplan}, {Kong}, {Kool}, {Mohite}, {Sollerman}, {Tzanidakis},
  {Webb}, \& {Zhang}}]{Andreoni+20}
{Andreoni}, I., {Goldstein}, D.~A., {Kasliwal}, M.~M., {et~al.} 2020,
  \href{http://dx.doi.org/10.3847/1538-4357/ab6a1b}{\JournalTitle{\apj}, 890,
  131}

\bibitem[{Antier {et~al.}(2020)}]{Antier:2020nuy}
Antier, S., {et~al.} 2020,
  \href{http://dx.doi.org/10.1093/mnras/staa1846}{\JournalTitle{Mon. Not. Roy.
  Astron. Soc.}, 497, 5518}

\bibitem[{{Ascenzi} {et~al.}(2019){Ascenzi}, {Coughlin}, {Dietrich}, {Foley},
  {Ramirez-Ruiz}, {Piranomonte}, {Mockler}, {Murguia-Berthier}, {Fryer},
  {Lloyd-Ronning}, \& {Rosswog}}]{2019MNRAS.486..672A}
{Ascenzi}, S., {Coughlin}, M.~W., {Dietrich}, T., {et~al.} 2019,
  \href{http://dx.doi.org/10.1093/mnras/stz891}{\JournalTitle{\mnras}, 486,
  672}

\bibitem[{{Berger}(2014)}]{Berger14}
{Berger}, E. 2014,
  \href{http://dx.doi.org/10.1146/annurev-astro-081913-035926}{\JournalTitle{\araa},
  52, 43}

\bibitem[{{Burns} {et~al.}(2016){Burns}, {Connaughton}, {Zhang}, {Lien},
  {Briggs}, {Goldstein}, {Pelassa}, \& {Troja}}]{Burns+16}
{Burns}, E., {Connaughton}, V., {Zhang}, B.-B., {et~al.} 2016,
  \href{http://dx.doi.org/10.3847/0004-637X/818/2/110}{\JournalTitle{\apj},
  818, 110}

\bibitem[{{Cai} \& {Yang}(2017)}]{2017PhRvD..95d4024C}
{Cai}, R.-G., \& {Yang}, T. 2017,
  \href{http://dx.doi.org/10.1103/PhysRevD.95.044024}{\JournalTitle{\prd}, 95,
  044024}

\bibitem[{{Camp} \& {TAP Team}(2019)}]{Camp19}
{Camp}, J., \& {TAP Team}. 2019, in The Space Astrophysics Landscape for the
  2020s and Beyond, Vol. 2135, 5027

\bibitem[{{Camp} {et~al.}(2019){Camp}, {Abel}, {Barthelmy}, {Bautz}, {Behar},
  {Berger}, {Spolaor}, {Cenko}, {Cornish}, {Dal Canton}, {Fryer}, {Gezari},
  {Gorenstein}, {Guiriec}, {Hartmann}, {Kalogera}, {Kouveliotou}, {Kruk},
  {Kutyrev}, {Margutti}, {Marshall}, {Metzger}, {Miller}, {Noble}, {Perkins},
  {Ptak}, {Purcell}, {Racusin}, {Schlieder}, {Schnittman}, {Sesana}, {Shawhan},
  {Singer}, {van der Horst}, {Willingale}, {Wood}, \& {Zhang}}]{TAP+19}
{Camp}, J., {Abel}, J., {Barthelmy}, S., {et~al.} 2019, in \baas, Vol.~51, 85

\bibitem[{{Chen}(2020)}]{2020arXiv200602779C}
{Chen}, H.-Y. 2020, \JournalTitle{arXiv e-prints}, arXiv:2006.02779

\bibitem[{{Chen} {et~al.}(2018){Chen}, {Fishbach}, \&
  {Holz}}]{2018Natur.562..545C}
{Chen}, H.-Y., {Fishbach}, M., \& {Holz}, D.~E. 2018,
  \href{http://dx.doi.org/10.1038/s41586-018-0606-0}{\JournalTitle{\nat}, 562,
  545}

\bibitem[{{Chen} {et~al.}(2020){Chen}, {Haster}, {Vitale}, {Farr}, \&
  {Isi}}]{2020arXiv200914057C}
{Chen}, H.-Y., {Haster}, C.-J., {Vitale}, S., {Farr}, W.~M., \& {Isi}, M. 2020,
  \JournalTitle{arXiv e-prints}, arXiv:2009.14057

\bibitem[{{Chen} \& {Holz}(2016{\natexlab{a}})}]{Chen&Holz16}
{Chen}, H.-Y., \& {Holz}, D.~E. 2016{\natexlab{a}}, \JournalTitle{arXiv
  e-prints}, arXiv:1612.01471

\bibitem[{{Chen} \& {Holz}(2016{\natexlab{b}})}]{loc3d}
---. 2016{\natexlab{b}}, \JournalTitle{arXiv e-prints}, arXiv:1612.01471

\bibitem[{{Chen} \& {Holz}(2017)}]{loc2d}
---. 2017,
  \href{http://dx.doi.org/10.3847/1538-4357/aa6f0d}{\JournalTitle{\apj}, 840,
  88}

\bibitem[{{Chen} {et~al.}(2017){Chen}, {Holz}, {Miller}, {Evans}, {Vitale}, \&
  {Creighton}}]{Chen+17}
{Chen}, H.-Y., {Holz}, D.~E., {Miller}, J., {et~al.} 2017, \JournalTitle{arXiv
  e-prints}, arXiv:1709.08079

\bibitem[{{Chen} {et~al.}(2019){Chen}, {Vitale}, \&
  {Narayan}}]{2019PhRvX...9c1028C}
{Chen}, H.-Y., {Vitale}, S., \& {Narayan}, R. 2019,
  \href{http://dx.doi.org/10.1103/PhysRevX.9.031028}{\JournalTitle{Physical
  Review X}, 9, 031028}

\bibitem[{{Cordier} {et~al.}(2015){Cordier}, {Wei}, {Atteia}, {Basa}, {Claret},
  {Daigne}, {Deng}, {Dong}, {Godet}, {Goldwurm}, {G{\"o}tz}, {Han}, {Klotz},
  {Lachaud}, {Osborne}, {Qiu}, {Schanne}, {Wu}, {Wang}, {Wu}, {Xin}, {Zhang},
  \& {Zhang}}]{Cordier+15}
{Cordier}, B., {Wei}, J., {Atteia}, J.~L., {et~al.} 2015, \JournalTitle{arXiv
  e-prints}, arXiv:1512.03323

\bibitem[{{Coulter} {et~al.}(2017){Coulter}, {Foley}, {Kilpatrick}, {Drout},
  {Piro}, {Shappee}, {Siebert}, {Simon}, {Ulloa}, {Kasen}, {Madore},
  {Murguia-Berthier}, {Pan}, {Prochaska}, {Ramirez-Ruiz}, {Rest}, \&
  {Rojas-Bravo}}]{2017Sci...358.1556C}
{Coulter}, D.~A., {Foley}, R.~J., {Kilpatrick}, C.~D., {et~al.} 2017,
  \href{http://dx.doi.org/10.1126/science.aap9811}{\JournalTitle{Science}, 358,
  1556}

\bibitem[{{Cowperthwaite} \& {Berger}(2015)}]{Cowperthwaite&Berger15}
{Cowperthwaite}, P.~S., \& {Berger}, E. 2015,
  \href{http://dx.doi.org/10.1088/0004-637X/814/1/25}{\JournalTitle{\apj}, 814,
  25}

\bibitem[{{Cowperthwaite} {et~al.}(2019{\natexlab{a}}){Cowperthwaite}, {Chen},
  {Margalit}, {Margutti}, {May}, {Metzger}, \& {Pankow}}]{Cowperthwaite+19b}
{Cowperthwaite}, P.~S., {Chen}, H.-Y., {Margalit}, B., {et~al.}
  2019{\natexlab{a}}, \JournalTitle{arXiv e-prints}, arXiv:1904.02718

\bibitem[{{Cowperthwaite} {et~al.}(2019{\natexlab{b}}){Cowperthwaite},
  {Villar}, {Scolnic}, \& {Berger}}]{Cowperthwaite+19}
{Cowperthwaite}, P.~S., {Villar}, V.~A., {Scolnic}, D.~M., \& {Berger}, E.
  2019{\natexlab{b}},
  \href{http://dx.doi.org/10.3847/1538-4357/ab07b6}{\JournalTitle{\apj}, 874,
  88}

\bibitem[{{Cowperthwaite} {et~al.}(2017){Cowperthwaite}, {Berger}, {Villar},
  {Metzger}, {Nicholl}, {Chornock}, {Blanchard}, {Fong}, {Margutti},
  {Soares-Santos}, {Alexander}, {Allam}, {Annis}, {Brout}, {Brown}, {Butler},
  {Chen}, {Diehl}, {Doctor}, {Drout}, {Eftekhari}, {Farr}, {Finley}, {Foley},
  {Frieman}, {Fryer}, {Garc{\'\i}a-Bellido}, {Gill}, {Guillochon}, {Herner},
  {Holz}, {Kasen}, {Kessler}, {Marriner}, {Matheson}, {Neilsen}, {Quataert},
  {Palmese}, {Rest}, {Sako}, {Scolnic}, {Smith}, {Tucker}, {Williams},
  {Balbinot}, {Carlin}, {Cook}, {Durret}, {Li}, {Lopes}, {Louren{\c{c}}o},
  {Marshall}, {Medina}, {Muir}, {Mu{\~n}oz}, {Sauseda}, {Schlegel}, {Secco},
  {Vivas}, {Wester}, {Zenteno}, {Zhang}, {Abbott}, {Banerji}, {Bechtol},
  {Benoit-L{\'e}vy}, {Bertin}, {Buckley-Geer}, {Burke}, {Capozzi}, {Carnero
  Rosell}, {Carrasco Kind}, {Castander}, {Crocce}, {Cunha}, {D'Andrea}, {da
  Costa}, {Davis}, {DePoy}, {Desai}, {Dietrich}, {Drlica-Wagner}, {Eifler},
  {Evrard}, {Fernand ez}, {Flaugher}, {Fosalba}, {Gaztanaga}, {Gerdes},
  {Giannantonio}, {Goldstein}, {Gruen}, {Gruendl}, {Gutierrez}, {Honscheid},
  {Jain}, {James}, {Jeltema}, {Johnson}, {Johnson}, {Kent}, {Krause}, {Kron},
  {Kuehn}, {Nuropatkin}, {Lahav}, {Lima}, {Lin}, {Maia}, {March}, {Martini},
  {McMahon}, {Menanteau}, {Miller}, {Miquel}, {Mohr}, {Neilsen}, {Nichol},
  {Ogando}, {Plazas}, {Roe}, {Romer}, {Roodman}, {Rykoff}, {Sanchez},
  {Scarpine}, {Schindler}, {Schubnell}, {Sevilla-Noarbe}, {Smith}, {Smith},
  {Sobreira}, {Suchyta}, {Swanson}, {Tarle}, {Thomas}, {Thomas}, {Troxel},
  {Vikram}, {Walker}, {Wechsler}, {Weller}, {Yanny}, \&
  {Zuntz}}]{Cowperthwaite+17}
{Cowperthwaite}, P.~S., {Berger}, E., {Villar}, V.~A., {et~al.} 2017,
  \href{http://dx.doi.org/10.3847/2041-8213/aa8fc7}{\JournalTitle{\apjl}, 848,
  L17}

\bibitem[{{Darbha} \& {Kasen}(2020)}]{Darbha&Kasen20}
{Darbha}, S., \& {Kasen}, D. 2020, \JournalTitle{arXiv e-prints},
  arXiv:2002.00299

\bibitem[{{DESI Collaboration} {et~al.}(2016){DESI Collaboration}, {Aghamousa},
  {et~al.}}]{DESI+16}
{DESI Collaboration}, {Aghamousa}, A., {et~al.} 2016, \JournalTitle{arXiv
  e-prints}, arXiv:1611.00036

\bibitem[{{Dore} {et~al.}(2019){Dore}, {Hirata}, {Wang}, {Weinberg}, {Eifler},
  {Foley}, {Heinrich}, {Krause}, {Perlmutter}, {Pisani}, {Scolnic}, {Spergel},
  {Suntzeff}, {Aldering}, {Baltay}, {Capak}, {Choi}, {Dvorkin}, {Fall}, {Fang},
  {Fruchter}, {Galbany}, {Ho}, {Hounsell}, {Izard}, {Jain}, {Koekemoer},
  {Kruk}, {Leauthaud}, {Malhotra}, {Mandelbaum}, {Massara}, {Masters},
  {Miyatake}, {Plazas}, {Rhoads}, {Rhodes}, {Rose}, {Rubin}, {Sako},
  {Samushia}, {Shirasaki}, {Simet}, {Takada}, {Troxel}, {Wu}, {Yoshida}, \&
  {Zhai}}]{Dore+19}
{Dore}, O., {Hirata}, C., {Wang}, Y., {et~al.} 2019, \JournalTitle{\baas}, 51,
  341

\bibitem[{{Farr} {et~al.}(2019){Farr}, {Fishbach}, {Ye}, \&
  {Holz}}]{2019ApJ...883L..42F}
{Farr}, W.~M., {Fishbach}, M., {Ye}, J., \& {Holz}, D.~E. 2019,
  \href{http://dx.doi.org/10.3847/2041-8213/ab4284}{\JournalTitle{\apjl}, 883,
  L42}

\bibitem[{{Feeney} {et~al.}(2019){Feeney}, {Peiris}, {Williamson}, {Nissanke},
  {Mortlock}, {Alsing}, \& {Scolnic}}]{Feeney+19}
{Feeney}, S.~M., {Peiris}, H.~V., {Williamson}, A.~R., {et~al.} 2019,
  \href{http://dx.doi.org/10.1103/PhysRevLett.122.061105}{\JournalTitle{\prl},
  122, 061105}

\bibitem[{{Fishbach} {et~al.}(2019){Fishbach}, {Gray}, {Maga{\~n}a Hernandez},
  {Qi}, {Sur}, {Acernese}, {Aiello}, {Allocca}, {Aloy}, {Amato}, \&
  et~al.}]{Fishbach+19}
{Fishbach}, M., {Gray}, R., {Maga{\~n}a Hernandez}, I., {et~al.} 2019,
  \href{http://dx.doi.org/10.3847/2041-8213/aaf96e}{\JournalTitle{\apjl}, 871,
  L13}

\bibitem[{{Foley} {et~al.}(2019){Foley}, {Bloom}, {Cenko}, {Chornock},
  {Dimitriadis}, {Dor{\'e}}, {Filippenko}, {Fox}, {Hirata}, {Jha}, {Jones},
  {Kasliwal}, {Kelly}, {Kilpatrick}, {Kirshner}, {Koekemoer}, {Kruk}, {Mandel},
  {Margutti}, {Miranda}, {Nissanke}, {Rest}, {Rhodes}, {Rodney}, {Rose},
  {Sand}, {Scolnic}, {Siellez}, {Smith}, {Spergel}, {Strolger}, {Suntzeff},
  {Wang}, \& {Wollack}}]{Foley+19}
{Foley}, R., {Bloom}, J.~S., {Cenko}, S.~B., {et~al.} 2019,
  \JournalTitle{\baas}, 51, 305

\bibitem[{{Fong} {et~al.}(2017){Fong}, {Berger}, {Blanchard}, {Margutti},
  {Cowperthwaite}, {Chornock}, {Alexander}, {Metzger}, {Villar}, {Nicholl},
  {Eftekhari}, {Williams}, {Annis}, {Brout}, {Brown}, {Chen}, {Doctor},
  {Diehl}, {Holz}, {Rest}, {Sako}, \& {Soares-Santos}}]{Fong+17}
{Fong}, W., {Berger}, E., {Blanchard}, P.~K., {et~al.} 2017,
  \href{http://dx.doi.org/10.3847/2041-8213/aa9018}{\JournalTitle{\apjl}, 848,
  L23}

\bibitem[{{Fong} {et~al.}(2020){Fong}, {Laskar}, {Rastinejad}, {Rouco
  Escorial}, {Schroeder}, {Barnes}, {Kilpatrick}, {Paterson}, {Berger},
  {Metzger}, {Dong}, {Nugent}, {Strausbaugh}, {Blanchard}, {Goyal},
  {Cucchiara}, {Terreran}, {Alexander}, {Eftekhari}, {Fryer}, {Margalit},
  {Margutti}, \& {Nicholl}}]{Fong+20}
{Fong}, W., {Laskar}, T., {Rastinejad}, J., {et~al.} 2020, \JournalTitle{arXiv
  e-prints}, arXiv:2008.08593

\bibitem[{{Fontes} {et~al.}(2020){Fontes}, {Fryer}, {Hungerford}, {Wollaeger},
  \& {Korobkin}}]{Fontes+20}
{Fontes}, C.~J., {Fryer}, C.~L., {Hungerford}, A.~L., {Wollaeger}, R.~T., \&
  {Korobkin}, O. 2020,
  \href{http://dx.doi.org/10.1093/mnras/staa485}{\JournalTitle{\mnras}, 493,
  4143}

\bibitem[{{Foreman-Mackey} {et~al.}(2013){Foreman-Mackey}, {Hogg}, {Lang}, \&
  {Goodman}}]{2013PASP..125..306F}
{Foreman-Mackey}, D., {Hogg}, D.~W., {Lang}, D., \& {Goodman}, J. 2013,
  \href{http://dx.doi.org/10.1086/670067}{\JournalTitle{\pasp}, 125, 306}

\bibitem[{{Freedman} {et~al.}(2019){Freedman}, {Madore}, {Hatt}, {Hoyt},
  {Jang}, {Beaton}, {Burns}, {Lee}, {Monson}, {Neeley}, {Phillips}, {Rich}, \&
  {Seibert}}]{Freedman+19}
{Freedman}, W.~L., {Madore}, B.~F., {Hatt}, D., {et~al.} 2019,
  \href{http://dx.doi.org/10.3847/1538-4357/ab2f73}{\JournalTitle{\apj}, 882,
  34}

\bibitem[{{Goldstein} {et~al.}(2017){Goldstein}, {Veres}, {Burns}, {Briggs},
  {Hamburg}, {Kocevski}, {Wilson-Hodge}, {Preece}, {Poolakkil}, {Roberts},
  {Hui}, {Connaughton}, {Racusin}, {von Kienlin}, {Dal Canton}, {Christensen},
  {Littenberg}, {Siellez}, {Blackburn}, {Broida}, {Bissaldi}, {Cleveland},
  {Gibby}, {Giles}, {Kippen}, {McBreen}, {McEnery}, {Meegan}, {Paciesas}, \&
  {Stanbro}}]{Goldstein+17}
{Goldstein}, A., {Veres}, P., {Burns}, E., {et~al.} 2017,
  \href{http://dx.doi.org/10.3847/2041-8213/aa8f41}{\JournalTitle{\apjl}, 848,
  L14}

\bibitem[{{Gomez} {et~al.}(2019){Gomez}, {Hosseinzadeh}, {Cowperthwaite},
  {Villar}, {Berger}, {Gardner}, {Alexand er}, {Blanchard}, {Chornock},
  {Drout}, {Eftekhari}, {Fong}, {Gill}, {Margutti}, {Nicholl}, {Paterson}, \&
  {Williams}}]{Gomez+19}
{Gomez}, S., {Hosseinzadeh}, G., {Cowperthwaite}, P.~S., {et~al.} 2019,
  \href{http://dx.doi.org/10.3847/2041-8213/ab4ad5}{\JournalTitle{\apjl}, 884,
  L55}

\bibitem[{{Gompertz} {et~al.}(2018){Gompertz}, {Levan}, {Tanvir}, {Hjorth},
  {Covino}, {Evans}, {Fruchter}, {Gonz{\'a}lez-Fern{\'a}ndez}, {Jin}, {Lyman},
  {Oates}, {O'Brien}, \& {Wiersema}}]{Gompertz+18}
{Gompertz}, B.~P., {Levan}, A.~J., {Tanvir}, N.~R., {et~al.} 2018,
  \href{http://dx.doi.org/10.3847/1538-4357/aac206}{\JournalTitle{\apj}, 860,
  62}

\bibitem[{{Guidorzi} {et~al.}(2017){Guidorzi}, {Margutti}, {Brout}, {Scolnic},
  {Fong}, {Alexander}, {Cowperthwaite}, {Annis}, {Berger}, {Blanchard},
  {Chornock}, {Coppejans}, {Eftekhari}, {Frieman}, {Huterer}, {Nicholl},
  {Soares-Santos}, {Terreran}, {Villar}, \& {Williams}}]{Guidorzi+17}
{Guidorzi}, C., {Margutti}, R., {Brout}, D., {et~al.} 2017,
  \href{http://dx.doi.org/10.3847/2041-8213/aaa009}{\JournalTitle{\apjl}, 851,
  L36}

\bibitem[{{Holz} \& {Hughes}(2005)}]{HolzHughes05}
{Holz}, D.~E., \& {Hughes}, S.~A. 2005,
  \href{http://dx.doi.org/10.1086/431341}{\JournalTitle{\apj}, 629, 15}

\bibitem[{{Hosseinzadeh} {et~al.}(2019){Hosseinzadeh}, {Cowperthwaite},
  {Gomez}, {Villar}, {Nicholl}, {Margutti}, {Berger}, {Chornock}, {Paterson},
  {Fong}, {Savchenko}, {Short}, {Alexander}, {Blanchard}, {Braga}, {Calkins},
  {Cartier}, {Coppejans}, {Eftekhari}, {Laskar}, {Ly}, {Patton}, {Pelisoli},
  {Reichart}, {Terreran}, \& {Williams}}]{Hosseinzadeh+19}
{Hosseinzadeh}, G., {Cowperthwaite}, P.~S., {Gomez}, S., {et~al.} 2019,
  \href{http://dx.doi.org/10.3847/2041-8213/ab271c}{\JournalTitle{\apjl}, 880,
  L4}

\bibitem[{{Husa} {et~al.}(2016){Husa}, {Khan}, {Hannam}, {P{\"u}rrer}, {Ohme},
  {Forteza}, \& {Boh{\'e}}}]{2016PhRvD..93d4006H}
{Husa}, S., {Khan}, S., {Hannam}, M., {et~al.} 2016,
  \href{http://dx.doi.org/10.1103/PhysRevD.93.044006}{\JournalTitle{\prd}, 93,
  044006}

\bibitem[{{Ivezi{\'c}} {et~al.}(2019){Ivezi{\'c}}, {Kahn},
  {et~al.}}]{Ivezic+19}
{Ivezi{\'c}}, {\v{Z}}., {Kahn}, S.~M., {et~al.} 2019,
  \href{http://dx.doi.org/10.3847/1538-4357/ab042c}{\JournalTitle{\apj}, 873,
  111}

\bibitem[{{Jin} {et~al.}(2020){Jin}, {He}, {Xu}, {Zhang}, \&
  {Zhang}}]{2020JCAP...03..051J}
{Jin}, S.-J., {He}, D.-Z., {Xu}, Y., {Zhang}, J.-F., \& {Zhang}, X. 2020,
  \href{http://dx.doi.org/10.1088/1475-7516/2020/03/051}{\JournalTitle{\jcap},
  2020, 051}

\bibitem[{{Kagra Collaboration} {et~al.}(2019){Kagra Collaboration}, {Akutsu},
  {Ando}, {Arai}, {Arai}, {Araki}, {Araya}, {Aritomi}, {Asada}, {Aso},
  {Atsuta}, {Awai}, {Bae}, {Baiotti}, {Barton}, {Cannon}, {Capocasa}, {Chen},
  {Chiu}, {Cho}, {Chu}, {Craig}, {Creus}, {Doi}, {Eda}, {Enomoto}, {Flaminio},
  {Fujii}, {Fujimoto}, {Fukunaga}, {Fukushima}, {Furuhata}, {Haino},
  {Hasegawa}, {Hashino}, {Hayama}, {Hirobayashi}, {Hirose}, {Hsieh}, {Huang},
  {Ikenoue}, {Inoue}, {Ioka}, {Itoh}, {Izumi}, {Kaji}, {Kajita}, {Kakizaki},
  {Kamiizumi}, {Kanbara}, {Kanda}, {Kanemura}, {Kaneyama}, {Kang}, {Kasuya},
  {Kataoka}, {Kawai}, {Kawamura}, {Kawasaki}, {Kim}, {Kim}, {Kim}, {Kim},
  {Kim}, {Kimura}, {Kinugawa}, {Kirii}, {Kitaoka}, {Kitazawa}, {Kojima},
  {Kokeyama}, {Komori}, {Kong}, {Kotake}, {Kozu}, {Kumar}, {Kuo}, {Kuroyanagi},
  {Lee}, {Lee}, {Lee}, {Leonardi}, {Lin}, {Lin}, {Liu}, {Liu}, {Majorana},
  {Mano}, {Marchio}, {Matsui}, {Matsushima}, {Michimura}, {Mio}, {Miyakawa},
  {Miyamoto}, {Miyamoto}, {Miyo}, {Miyoki}, {Morii}, {Morisaki}, {Moriwaki},
  {Morozumi}, {Musha}, {Nagano}, {Nagano}, {Nakamura}, {Nakamura}, {Nakano},
  {Nakano}, {Nakao}, {Narikawa}, {Naticchioni}, {Nguyen Quynh}, {Ni},
  {Nishizawa}, {Obuchi}, {Ochi}, {Oh}, {Oh}, {Ohashi}, {Ohishi}, {Ohkawa},
  {Okutomi}, {Ono}, {Oohara}, {Ooi}, {Pan}, {Park}, {Pe{\~n}a Arellano},
  {Pinto}, {Sago}, {Saijo}, {Saitou}, {Saito}, {Sakai}, {Sakai}, {Sakai},
  {Sasai}, {Sasaki}, {Sasaki}, {Sato}, {Sato}, {Sato}, {Sekiguchi}, {Seto},
  {Shibata}, {Shimoda}, {Shinkai}, {Shishido}, {Shoda}, {Somiya}, {Son},
  {Suemasa}, {Suzuki}, {Suzuki}, {Tagoshi}, {Tahara}, {Takahashi}, {Takahashi},
  {Takamori}, {Takeda}, {Tanaka}, {Tanaka}, {Tanaka}, {Tanioka}, {Tapia San
  Martin}, {Tatsumi}, {Tomaru}, {Tomura}, {Travasso}, {Tsubono}, {Tsuchida},
  {Uchikata}, {Uchiyama}, {Uehara}, {Ueki}, {Ueno}, {Uraguchi}, {Ushiba}, {van
  Putten}, {Vocca}, {Wada}, {Wakamatsu}, {Watanabe}, {Xu}, {Yamada},
  {Yamamoto}, {Yamamoto}, {Yamamoto}, {Yamamoto}, {Yamamoto}, {Yokogawa},
  {Yokoyama}, {Yokozawa}, {Yoon}, {Yoshioka}, {Yuzurihara}, {Zeidler}, \&
  {Zhu}}]{KAGRA+19}
{Kagra Collaboration}, {Akutsu}, T., {Ando}, M., {et~al.} 2019,
  \href{http://dx.doi.org/10.1038/s41550-018-0658-y}{\JournalTitle{Nature
  Astronomy}, 3, 35}

\bibitem[{{Kasen} {et~al.}(2015){Kasen}, {Fern{\'a}ndez}, \&
  {Metzger}}]{Kasen+15}
{Kasen}, D., {Fern{\'a}ndez}, R., \& {Metzger}, B.~D. 2015,
  \href{http://dx.doi.org/10.1093/mnras/stv721}{\JournalTitle{\mnras}, 450,
  1777}

\bibitem[{{Khan} {et~al.}(2016){Khan}, {Husa}, {Hannam}, {Ohme}, {P{\"u}rrer},
  {Forteza}, \& {Boh{\'e}}}]{2016PhRvD..93d4007K}
{Khan}, S., {Husa}, S., {Hannam}, M., {et~al.} 2016,
  \href{http://dx.doi.org/10.1103/PhysRevD.93.044007}{\JournalTitle{\prd}, 93,
  044007}

\bibitem[{{Knox} \& {Millea}(2019)}]{Knox&Millea19}
{Knox}, L., \& {Millea}, M. 2019, \JournalTitle{arXiv e-prints},
  arXiv:1908.03663

\bibitem[{{Madau} \& {Dickinson}(2014)}]{Madau&Dickinson14}
{Madau}, P., \& {Dickinson}, M. 2014,
  \href{http://dx.doi.org/10.1146/annurev-astro-081811-125615}{\JournalTitle{\araa},
  52, 415}

\bibitem[{{Mandel} {et~al.}(2019){Mandel}, {Farr}, \&
  {Gair}}]{2019MNRAS.486.1086M}
{Mandel}, I., {Farr}, W.~M., \& {Gair}, J.~R. 2019,
  \href{http://dx.doi.org/10.1093/mnras/stz896}{\JournalTitle{\mnras}, 486,
  1086}

\bibitem[{{Margalit} \& {Metzger}(2017)}]{Margalit&Metzger17}
{Margalit}, B., \& {Metzger}, B.~D. 2017,
  \href{http://dx.doi.org/10.3847/2041-8213/aa991c}{\JournalTitle{\apjl}, 850,
  L19}

\bibitem[{{Margalit} \& {Metzger}(2019)}]{Margalit&Metzger19}
---. 2019,
  \href{http://dx.doi.org/10.3847/2041-8213/ab2ae2}{\JournalTitle{\apjl}, 880,
  L15}

\bibitem[{{Margutti} {et~al.}(2017){Margutti}, {Berger}, {Fong}, {Guidorzi},
  {Alexander}, {Metzger}, {Blanchard}, {Cowperthwaite}, {Chornock},
  {Eftekhari}, {Nicholl}, {Villar}, {Williams}, {Annis}, {Brown}, {Chen},
  {Doctor}, {Frieman}, {Holz}, {Sako}, \& {Soares-Santos}}]{Margutti+17}
{Margutti}, R., {Berger}, E., {Fong}, W., {et~al.} 2017,
  \href{http://dx.doi.org/10.3847/2041-8213/aa9057}{\JournalTitle{\apjl}, 848,
  L20}

\bibitem[{{Margutti} {et~al.}(2018{\natexlab{a}}){Margutti}, {Cowperthwaite},
  {Doctor}, {Mortensen}, {Pankow}, {Salafia}, {Villar}, {Alexander}, {Annis},
  {Andreoni}, {Baldeschi}, {Balmaverde}, {Berger}, {Bernardini}, {Berry},
  {Bianco}, {Blanchard}, {Brocato}, {Carnerero}, {Cartier}, {Cenko},
  {Chornock}, {Chomiuk}, {Copperwheat}, {Coughlin}, {Coppejans}, {Corsi},
  {D'Ammando}, {Datrier}, {D'Avanzo}, {Dimitriadis}, {Drout}, {Foley}, {Fong},
  {Fox}, {Ghirlanda}, {Goldstein}, {Grindlay}, {Guidorzi}, {Haiman}, {Hendry},
  {Holz}, {Hung}, {Inserra}, {Jones}, {Kalogera}, {Kilpatrick}, {Lamb},
  {Laskar}, {Levan}, {Mason}, {Maguire}, {Melandri}, {Milisavljevic}, {Miller},
  {Narayan}, {Nielsen}, {Nicholl}, {Nissanke}, {Nugent}, {Pan}, {Pasham},
  {Paterson}, {Piranomonte}, {Racusin}, {Rest}, {Righi}, {Sand}, {Seaman},
  {Scolnic}, {Siellez}, {Singer}, {Szkody}, {Smith}, {Steeghs}, {Sullivan},
  {Tanvir}, {Terreran}, {Trimble}, {Valenti}, {LSST Transient}, \& {Variable
  Stars Collaboration}}]{Margutti+18}
{Margutti}, R., {Cowperthwaite}, P., {Doctor}, Z., {et~al.} 2018{\natexlab{a}},
  \JournalTitle{arXiv e-prints}, arXiv:1812.04051

\bibitem[{{Margutti} {et~al.}(2018{\natexlab{b}}){Margutti}, {Cowperthwaite},
  {Doctor}, {Mortensen}, {Pankow}, {Salafia}, {Villar}, {Alexander}, {Annis},
  {Andreoni}, {Baldeschi}, {Balmaverde}, {Berger}, {Bernardini}, {Berry},
  {Bianco}, {Blanchard}, {Brocato}, {Carnerero}, {Cartier}, {Cenko},
  {Chornock}, {Chomiuk}, {Copperwheat}, {Coughlin}, {Coppejans}, {Corsi},
  {D'Ammando}, {Datrier}, {D'Avanzo}, {Dimitriadis}, {Drout}, {Foley}, {Fong},
  {Fox}, {Ghirlanda}, {Goldstein}, {Grindlay}, {Guidorzi}, {Haiman}, {Hendry},
  {Holz}, {Hung}, {Inserra}, {Jones}, {Kalogera}, {Kilpatrick}, {Lamb},
  {Laskar}, {Levan}, {Mason}, {Maguire}, {Melandri}, {Milisavljevic}, {Miller},
  {Narayan}, {Nielsen}, {Nicholl}, {Nissanke}, {Nugent}, {Pan}, {Pasham},
  {Paterson}, {Piranomonte}, {Racusin}, {Rest}, {Righi}, {Sand}, {Seaman},
  {Scolnic}, {Siellez}, {Singer}, {Szkody}, {Smith}, {Steeghs}, {Sullivan},
  {Tanvir}, {Terreran}, {Trimble}, {Valenti}, {LSST Transient}, \& {Variable
  Stars Collaboration}}]{Cowperthwaite+18}
---. 2018{\natexlab{b}}, \JournalTitle{arXiv e-prints}, arXiv:1812.04051

\bibitem[{{McEnery} {et~al.}(2019){McEnery}, {van der Horst}, {Dominguez},
  {Moiseev}, {Marcowith}, {Harding}, {Lien}, {Giuliani}, {Inglis}, {Ansoldi},
  {Stamerra}, {Manousakis}, {Strong}, {Bambi}, {Patricelli}, {Baring},
  {Barrio}, {Bastieri}, {Fields}, {Beacom}, {Beckmann}, {Bednarek}, {Rani},
  {Boggs}, {Bolotnikov}, {Cenko}, {Buckley}, {Grefenstette}, {Hui}, {Pittori},
  {Prescod-Weinstein}, {Shrader}, {Gouiffes}, {Kierans}, {Wilson-Hodge},
  {D'Ammando}, {Castro}, {Kocveski}, {Gasparrini}, {Thompson}, {Williams}, {De
  Angelis}, {Bernard}, {Digel}, {Morcuende}, {Charles}, {Bissaldi}, {Hays},
  {Ferrara}, {Bozzo}, {Grove}, {Wulf}, {Bottacini}, {Caroli}, {Kislat},
  {Oikonomou}, {Giordano}, {Longo}, {Fryer}, {Fukazawa}, {Georganopoulos}, {De
  Nolfo}, {Vianello}, {Kanbach}, {Younes}, {Blumer}, {Hartmann}, {Hernanz},
  {Takahashi}, {Li}, {Agudo}, {Moskalenko}, {Stumke}, {Grenier}, {Smith},
  {Rodi}, {Perkins}, {Gelfand}, {Holder}, {Knodlseder}, {Kopp}, {Lenain},
  {{\'A}lvarez}, {Metcalfe}, {Krizmanic}, {Stephen}, {Hewitt}, {Mitchell},
  {Harding}, {Tomsick}, {Racusin}, {Finke}, {Kargaltsev}, {Klimenko},
  {Krawczynski}, {Smith}, {Kubo}, {Di Venere}, {Marcotulli}, {Lommler},
  {Parker}, {Baldini}, {Foffano}, {Zampieri}, {Tibaldo}, {Petropoulou},
  {Ajello}, {Meyer}, {L{\'o}pez}, {McConnell}, {Boettcher}, {Cardillo},
  {Martinez}, {Kerr}, {Mazziotta}, {McEnery}, {Di Mauro}, {Wood}, {Meyer},
  {Briggs}, {De Becker}, {Lovellette}, {Doro}, {Sanchez-Conde}, {Moss},
  {Mizuno}, {Rib{\'o}}, {Nakazawa}, {Neilson}, {Auricchio}, {Omodei},
  {Oberlack}, {Ohno}, {Orland o}, {Otte}, {Coppi}, {Bloser}, {Zhang},
  {Laurent}, {Pohl}, {Prand ini}, {Shawhan}, {Caputo}, {Campana}, {Rando},
  {Woolf}, {Johnson}, {Mignani}, {Walter}, {Ojha}, {da Silva}, {Dietrich},
  {Funk}, {Zane}, {Anton}, {Buson}, {Cutini}, {Saz Parkinson}, {Schirato},
  {Griffin}, {Kaufmann}, {Stawarz}, {Ciprini}, {Del Sordo}, {Jones}, {Guiriec},
  {Tajima}, {Cheung}, {The}, {Venters}, {Porter}, {Linden}, {Barres}, {Paliya},
  {Bozhilov}, {Vestrand}, {Tatischeff}, {Chen}, {Wang}, {Tanaka}, {Uhm},
  {Zhang}, {Zimmer}, {Zoglauer}, \& {Wadiasingh}}]{McEnery+19}
{McEnery}, J., {van der Horst}, A., {Dominguez}, A., {et~al.} 2019, in \baas,
  Vol.~51, 245

\bibitem[{{Metzger} \& {Berger}(2012)}]{Metzger&Berger12}
{Metzger}, B.~D., \& {Berger}, E. 2012,
  \href{http://dx.doi.org/10.1088/0004-637X/746/1/48}{\JournalTitle{\apj}, 746,
  48}

\bibitem[{{Metzger} {et~al.}(2010){Metzger}, {Mart{\'\i}nez-Pinedo}, {Darbha},
  {Quataert}, {Arcones}, {Kasen}, {Thomas}, {Nugent}, {Panov}, \&
  {Zinner}}]{Metzger+10}
{Metzger}, B.~D., {Mart{\'\i}nez-Pinedo}, G., {Darbha}, S., {et~al.} 2010,
  \href{http://dx.doi.org/10.1111/j.1365-2966.2010.16864.x}{\JournalTitle{\mnras},
  406, 2650}

\bibitem[{{Nissanke} {et~al.}(2011){Nissanke}, {Sievers}, {Dalal}, \&
  {Holz}}]{Nissanke+11}
{Nissanke}, S., {Sievers}, J., {Dalal}, N., \& {Holz}, D. 2011,
  \href{http://dx.doi.org/10.1088/0004-637X/739/2/99}{\JournalTitle{\apj}, 739,
  99}

\bibitem[{{Pian} {et~al.}(2017){Pian}, {D'Avanzo}, {Benetti}, {Branchesi},
  {Brocato}, {Campana}, {Cappellaro}, {Covino}, {D'Elia}, {Fynbo}, {Getman},
  {Ghirlanda}, {Ghisellini}, {Grado}, {Greco}, {Hjorth}, {Kouveliotou},
  {Levan}, {Limatola}, {Malesani}, {Mazzali}, {Melandri}, {M{\o}ller},
  {Nicastro}, {Palazzi}, {Piranomonte}, {Rossi}, {Salafia}, {Selsing},
  {Stratta}, {Tanaka}, {Tanvir}, {Tomasella}, {Watson}, {Yang}, {Amati},
  {Antonelli}, {Ascenzi}, {Bernardini}, {Bo{\"e}r}, {Bufano}, {Bulgarelli},
  {Capaccioli}, {Casella}, {Castro-Tirado}, {Chassande-Mottin}, {Ciolfi},
  {Copperwheat}, {Dadina}, {De Cesare}, {di Paola}, {Fan}, {Gendre},
  {Giuffrida}, {Giunta}, {Hunt}, {Israel}, {Jin}, {Kasliwal}, {Klose}, {Lisi},
  {Longo}, {Maiorano}, {Mapelli}, {Masetti}, {Nava}, {Patricelli}, {Perley},
  {Pescalli}, {Piran}, {Possenti}, {Pulone}, {Razzano}, {Salvaterra},
  {Schipani}, {Spera}, {Stamerra}, {Stella}, {Tagliaferri}, {Testa}, {Troja},
  {Turatto}, {Vergani}, \& {Vergani}}]{2017Natur.551...67P}
{Pian}, E., {D'Avanzo}, P., {Benetti}, S., {et~al.} 2017,
  \href{http://dx.doi.org/10.1038/nature24298}{\JournalTitle{\nat}, 551, 67}

\bibitem[{{Planck Collaboration} {et~al.}(2016){Planck Collaboration}, {Ade},
  {Aghanim}, {Arnaud}, {Ashdown}, {Aumont}, {Baccigalupi}, {Banday},
  {Barreiro}, {Bartlett}, \& et~al.}]{Planck2016}
{Planck Collaboration}, {Ade}, P.~A.~R., {Aghanim}, N., {et~al.} 2016,
  \href{http://dx.doi.org/10.1051/0004-6361/201525830}{\JournalTitle{\aap},
  594, A13}

\bibitem[{{Planck Collaboration} {et~al.}(2018){Planck Collaboration},
  {Aghanim}, {Akrami}, {Ashdown}, {Aumont}, {Baccigalupi}, {Ballardini},
  {Banday}, {Barreiro}, {Bartolo}, {Basak}, {Battye}, {Benabed}, {Bernard},
  {Bersanelli}, {Bielewicz}, {Bock}, {Bond}, {Borrill}, {Bouchet}, {Boulanger},
  {Bucher}, {Burigana}, {Butler}, {Calabrese}, {Cardoso}, {Carron},
  {Challinor}, {Chiang}, {Chluba}, {Colombo}, {Combet}, {Contreras}, {Crill},
  {Cuttaia}, {de Bernardis}, {de Zotti}, {Delabrouille}, {Delouis}, {Di
  Valentino}, {Diego}, {Dor{\'e}}, {Douspis}, {Ducout}, {Dupac}, {Dusini},
  {Efstathiou}, {Elsner}, {En{\ss}lin}, {Eriksen}, {Fantaye}, {Farhang},
  {Fergusson}, {Fernandez-Cobos}, {Finelli}, {Forastieri}, {Frailis},
  {Franceschi}, {Frolov}, {Galeotta}, {Galli}, {Ganga}, {G{\'e}nova-Santos},
  {Gerbino}, {Ghosh}, {Gonz{\'a}lez-Nuevo}, {G{\'o}rski}, {Gratton},
  {Gruppuso}, {Gudmundsson}, {Hamann}, {Handley}, {Herranz}, {Hivon}, {Huang},
  {Jaffe}, {Jones}, {Karakci}, {Keih{\"a}nen}, {Keskitalo}, {Kiiveri}, {Kim},
  {Kisner}, {Knox}, {Krachmalnicoff}, {Kunz}, {Kurki-Suonio}, {Lagache},
  {Lamarre}, {Lasenby}, {Lattanzi}, {Lawrence}, {Le Jeune}, {Lemos},
  {Lesgourgues}, {Levrier}, {Lewis}, {Liguori}, {Lilje}, {Lilley}, {Lindholm},
  {L{\'o}pez-Caniego}, {Lubin}, {Ma}, {Mac{\'{\i}}as-P{\'e}rez}, {Maggio},
  {Maino}, {Mandolesi}, {Mangilli}, {Marcos-Caballero}, {Maris}, {Martin},
  {Martinelli}, {Mart{\'{\i}}nez-Gonz{\'a}lez}, {Matarrese}, {Mauri}, {McEwen},
  {Meinhold}, {Melchiorri}, {Mennella}, {Migliaccio}, {Millea}, {Mitra},
  {Miville-Desch{\^e}nes}, {Molinari}, {Montier}, {Morgante}, {Moss}, {Natoli},
  {N{\o}rgaard-Nielsen}, {Pagano}, {Paoletti}, {Partridge}, {Patanchon},
  {Peiris}, {Perrotta}, {Pettorino}, {Piacentini}, {Polastri}, {Polenta},
  {Puget}, {Rachen}, {Reinecke}, {Remazeilles}, {Renzi}, {Rocha}, {Rosset},
  {Roudier}, {Rubi{\~n}o-Mart{\'{\i}}n}, {Ruiz-Granados}, {Salvati}, {Sandri},
  {Savelainen}, {Scott}, {Shellard}, {Sirignano}, {Sirri}, {Spencer},
  {Sunyaev}, {Suur-Uski}, {Tauber}, {Tavagnacco}, {Tenti}, {Toffolatti},
  {Tomasi}, {Trombetti}, {Valenziano}, {Valiviita}, {Van Tent}, {Vibert},
  {Vielva}, {Villa}, {Vittorio}, {Wandelt}, {Wehus}, {White}, {White},
  {Zacchei}, \& {Zonca}}]{2018arXiv180706209P}
{Planck Collaboration}, {Aghanim}, N., {Akrami}, Y., {et~al.} 2018,
  \JournalTitle{ArXiv e-prints},
  \href{http://arxiv.org/abs/1807.06209}{{\sffamily arXiv:1807.06209}}

\bibitem[{{Riess}(2019)}]{Riess19}
{Riess}, A.~G. 2019,
  \href{http://dx.doi.org/10.1038/s42254-019-0137-0}{\JournalTitle{Nature
  Reviews Physics}, 2, 10}

\bibitem[{{Riess} {et~al.}(2019){Riess}, {Casertano}, {Yuan}, {Macri}, \&
  {Scolnic}}]{2019ApJ...876...85R}
{Riess}, A.~G., {Casertano}, S., {Yuan}, W., {Macri}, L.~M., \& {Scolnic}, D.
  2019, \href{http://dx.doi.org/10.3847/1538-4357/ab1422}{\JournalTitle{\apj},
  876, 85}

\bibitem[{{Rossi} {et~al.}(2020){Rossi}, {Stratta}, {Maiorano}, {Spighi},
  {Masetti}, {Palazzi}, {Gardini}, {Melandri}, {Nicastro}, {Pian}, {Branchesi},
  {Dadina}, {Testa}, {Brocato}, {Benetti}, {Ciolfi}, {Covino}, {D'Elia},
  {Grado}, {Izzo}, {Perego}, {Piranomonte}, {Salvaterra}, {Selsing},
  {Tomasella}, {Yang}, {Vergani}, {Amati}, \& {Stephen}}]{2020MNRAS.493.3379R}
{Rossi}, A., {Stratta}, G., {Maiorano}, E., {et~al.} 2020,
  \href{http://dx.doi.org/10.1093/mnras/staa479}{\JournalTitle{\mnras}, 493,
  3379}

\bibitem[{{Sathyaprakash} {et~al.}(2010){Sathyaprakash}, {Schutz}, \& {Van Den
  Broeck}}]{2010CQGra..27u5006S}
{Sathyaprakash}, B.~S., {Schutz}, B.~F., \& {Van Den Broeck}, C. 2010,
  \href{http://dx.doi.org/10.1088/0264-9381/27/21/215006}{\JournalTitle{Classical
  and Quantum Gravity}, 27, 215006}

\bibitem[{{Savchenko} {et~al.}(2017){Savchenko}, {Ferrigno}, {Kuulkers},
  {Bazzano}, {Bozzo}, {Brandt}, {Chenevez}, {Courvoisier}, {Diehl}, {Domingo},
  {Hanlon}, {Jourdain}, {von Kienlin}, {Laurent}, {Lebrun}, {Lutovinov},
  {Martin-Carrillo}, {Mereghetti}, {Natalucci}, {Rodi}, {Roques}, {Sunyaev}, \&
  {Ubertini}}]{Savchenko+17}
{Savchenko}, V., {Ferrigno}, C., {Kuulkers}, E., {et~al.} 2017,
  \href{http://dx.doi.org/10.3847/2041-8213/aa8f94}{\JournalTitle{\apjl}, 848,
  L15}

\bibitem[{{Schutz}(1986)}]{Schutz1986}
{Schutz}, B.~F. 1986,
  \href{http://dx.doi.org/10.1038/323310a0}{\JournalTitle{\nat}, 323, 310}

\bibitem[{{Soares-Santos} {et~al.}(2017){Soares-Santos}, {Holz}, {Annis},
  {Chornock}, {Herner}, {Berger}, {Brout}, {Chen}, {Kessler}, {Sako}, {Allam},
  {Tucker}, {Butler}, {Palmese}, {Doctor}, {Diehl}, {Frieman}, {Yanny}, {Lin},
  {Scolnic}, {Cowperthwaite}, {Neilsen}, {Marriner}, {Kuropatkin}, {Hartley},
  {Paz-Chinch{\'o}n}, {Alexander}, {Balbinot}, {Blanchard}, {Brown}, {Carlin},
  {Conselice}, {Cook}, {Drlica-Wagner}, {Drout}, {Durret}, {Eftekhari}, {Farr},
  {Finley}, {Foley}, {Fong}, {Fryer}, {Garc{\'{\i}}a-Bellido}, {Gill},
  {Gruendl}, {Hanna}, {Kasen}, {Li}, {Lopes}, {Louren{\c c}o}, {Margutti},
  {Marshall}, {Matheson}, {Medina}, {Metzger}, {Mu{\~n}oz}, {Muir}, {Nicholl},
  {Quataert}, {Rest}, {Sauseda}, {Schlegel}, {Secco}, {Sobreira}, {Stebbins},
  {Villar}, {Vivas}, {Walker}, {Wester}, {Williams}, {Zenteno}, {Zhang},
  {Abbott}, {Abdalla}, {Banerji}, {Bechtol}, {Benoit-L{\'e}vy}, {Bertin},
  {Brooks}, {Buckley-Geer}, {Burke}, {Carnero Rosell}, {Carrasco Kind},
  {Carretero}, {Castander}, {Crocce}, {Cunha}, {D'Andrea}, {da Costa}, {Davis},
  {Desai}, {Dietrich}, {Doel}, {Eifler}, {Fernandez}, {Flaugher}, {Fosalba},
  {Gaztanaga}, {Gerdes}, {Giannantonio}, {Goldstein}, {Gruen}, {Gschwend},
  {Gutierrez}, {Honscheid}, {Jain}, {James}, {Jeltema}, {Johnson}, {Johnson},
  {Kent}, {Krause}, {Kron}, {Kuehn}, {Kuhlmann}, {Lahav}, {Lima}, {Maia},
  {March}, {McMahon}, {Menanteau}, {Miquel}, {Mohr}, {Nichol}, {Nord},
  {Ogando}, {Petravick}, {Plazas}, {Romer}, {Roodman}, {Rykoff}, {Sanchez},
  {Scarpine}, {Schubnell}, {Sevilla-Noarbe}, {Smith}, {Smith}, {Suchyta},
  {Swanson}, {Tarle}, {Thomas}, {Thomas}, {Troxel}, {Vikram}, {Wechsler},
  {Weller}, {Dark Energy Survey}, \& {Dark Energy Camera GW-EM
  Collaboration}}]{GW170817DECam}
{Soares-Santos}, M., {Holz}, D.~E., {Annis}, J., {et~al.} 2017,
  \href{http://dx.doi.org/10.3847/2041-8213/aa9059}{\JournalTitle{\apjl}, 848,
  L16}

\bibitem[{{Soares-Santos} {et~al.}(2019){Soares-Santos}, {Palmese}, {Hartley},
  {Annis}, {Garcia-Bellido}, {Lahav}, {Doctor}, {Fishbach}, {Holz}, {Lin}, \&
  et~al.}]{SoaresSantos+19}
{Soares-Santos}, M., {Palmese}, A., {Hartley}, W., {et~al.} 2019,
  \href{http://dx.doi.org/10.3847/2041-8213/ab14f1}{\JournalTitle{\apjl}, 876,
  L7}

\bibitem[{{Sun} {et~al.}(2020){Sun}, {Goetz}, {Kissel}, {Betzwieser}, {Karki},
  {Viets}, {Wade}, {Bhattacharjee}, {Bossilkov}, {Covas}, {Datrier}, {Gray},
  {Kand hasamy}, {Lecoeuche}, {Mendell}, {Mistry}, {Payne}, {Savage},
  {Weinstein}, {Aston}, {Buikema}, {Cahillane}, {Driggers}, {Dwyer}, {Kumar},
  \& {Urban}}]{2020arXiv200502531S}
{Sun}, L., {Goetz}, E., {Kissel}, J.~S., {et~al.} 2020, \JournalTitle{arXiv
  e-prints}, arXiv:2005.02531

\bibitem[{{Vieira} {et~al.}(2020){Vieira}, {Ruan}, {Haggard}, {Drout}, {Nynka},
  {Boyce}, {Spekkens}, {Safi-Harb}, {Carlberg}, {Fern{\'a}ndez}, {Piro},
  {Afsariardchi}, \& {Moon}}]{Vierira+20}
{Vieira}, N., {Ruan}, J.~J., {Haggard}, D., {et~al.} 2020,
  \href{http://dx.doi.org/10.3847/1538-4357/ab917d}{\JournalTitle{\apj}, 895,
  96}

\bibitem[{{Villar} {et~al.}(2017){Villar}, {Berger}, {Metzger}, \&
  {Guillochon}}]{Villar+17a}
{Villar}, V.~A., {Berger}, E., {Metzger}, B.~D., \& {Guillochon}, J. 2017,
  \href{http://dx.doi.org/10.3847/1538-4357/aa8fcb}{\JournalTitle{\apj}, 849,
  70}

\bibitem[{{Wanderman} \& {Piran}(2015)}]{Wanderman&Piran15}
{Wanderman}, D., \& {Piran}, T. 2015,
  \href{http://dx.doi.org/10.1093/mnras/stv123}{\JournalTitle{\mnras}, 448,
  3026}

\bibitem[{{Wong} {et~al.}(2020){Wong}, {Suyu}, {Chen}, {Rusu}, {Millon},
  {Sluse}, {Bonvin}, {Fassnacht}, {Taubenberger}, {Auger}, {Birrer}, {Chan},
  {Courbin}, {Hilbert}, {Tihhonova}, {Treu}, {Agnello}, {Ding}, {Jee},
  {Komatsu}, {Shajib}, {Sonnenfeld}, {Bland ford}, {Koopmans}, {Marshall}, \&
  {Meylan}}]{Wong+20}
{Wong}, K.~C., {Suyu}, S.~H., {Chen}, G. C.~F., {et~al.} 2020,
  \href{http://dx.doi.org/10.1093/mnras/stz3094}{\JournalTitle{\mnras}},
  \href{http://arxiv.org/abs/1907.04869}{{\sffamily arXiv:1907.04869
  [astro-ph.CO]}}

\bibitem[{{Zhao} {et~al.}(2011){Zhao}, {van den Broeck}, {Baskaran}, \&
  {Li}}]{2011PhRvD..83b3005Z}
{Zhao}, W., {van den Broeck}, C., {Baskaran}, D., \& {Li}, T.~G.~F. 2011,
  \href{http://dx.doi.org/10.1103/PhysRevD.83.023005}{\JournalTitle{\prd}, 83,
  023005}

\end{thebibliography}

\end{document}